\documentclass[aps,prb,twocolumn,amsmath,amssymb,superscriptaddress,longbibliography]{revtex4-1} 

\usepackage{graphicx}
\usepackage{dcolumn}
\usepackage{bm}
\usepackage{color}
\usepackage{epstopdf}
\usepackage{amsmath,amsthm}
\usepackage{amsfonts}
\usepackage[breaklinks,colorlinks,bookmarks=false,citecolor=blue,linkcolor=red,urlcolor=blue]{hyperref}

\begin{document}

\title{Triplet excitations in the frustrated spin ladder Li$_2$Cu$_2$O(SO$_4$)$_2$}

\author{O. Vaccarelli}
\affiliation{Sorbonne Universit\'e, UMR CNRS 7590, Mus\'eum National d'Histoire Naturelle, 
IRD UMR 206, Institut de Min\'eralogie, de Physique des Mat\'eriaux, et de Cosmochimie (IMPMC), 
4 place Jussieu, F-75005 Paris, France}

\author{A. Honecker}
\affiliation{Laboratoire de Physique Th\'eorique et Mod\'elisation, CNRS UMR 8089, Universit\'e de 
Cergy-Pontoise, F-95302 Cergy-Pontoise Cedex, France}

\author{P. Giura}
\affiliation{Sorbonne Universit\'e, UMR CNRS 7590, Mus\'eum National d'Histoire Naturelle, 
IRD UMR 206, Institut de Min\'eralogie, de Physique des Mat\'eriaux, et de Cosmochimie (IMPMC), 
4 place Jussieu, F-75005 Paris, France}

\author{K. B\'eneut}
\affiliation{Sorbonne Universit\'e, UMR CNRS 7590, Mus\'eum National d'Histoire Naturelle, 
IRD UMR 206, Institut de Min\'eralogie, de Physique des Mat\'eriaux, et de Cosmochimie (IMPMC), 
4 place Jussieu, F-75005 Paris, France}

\author{B. F{\aa}k}
\affiliation{Institut Laue-Langevin, CS 20156, F-38042 Grenoble Cedex 9, France}

\author{G. Rousse}
\affiliation{Coll\`ege de France, Chimie du Solide et de l'Energie, UMR 8260, 11 place Marcelin Berthelot,  
F-75231 Paris Cedex 05, France}

\author{G. Radtke}
\email{guillaume.radtke@sorbonne-universite.fr}
\affiliation{Sorbonne Universit\'e, UMR CNRS 7590, Mus\'eum National d'Histoire Naturelle, 
IRD UMR 206, Institut de Min\'eralogie, de Physique des Mat\'eriaux, et de Cosmochimie (IMPMC), 
4 place Jussieu, F-75005 Paris, France}

\begin{abstract}
Magnetic excitations of the recently discovered frustrated spin-1/2  two-leg ladder system Li$_2$Cu$_2$O(SO$_4$)$_2$ 
are investigated using  inelastic neutron scattering, magnetic susceptibility and infrared absorption measurements. 
Despite the presence of a magnetic dimerization concomitant with the tetragonal-to-triclinic structural distortion 
occurring below 125~K, neutron scattering experiments reveal the presence of dispersive triplet excitations above 
a spin gap of $\Delta = 10.6 \text { meV}$ at $1.5 \text{ K}$, a value consistent with the estimates extracted 
from magnetic susceptibility. The likely detection of these spin excitations in infrared spectroscopy 
is explained by invoking a dynamic Dzyaloshinskii-Moriya mechanism in which light is coupled to the dimer 
singlet-to-triplet transition through an optical phonon. These results are qualitatively explained by exact diagonalization 
and higher-order perturbation calculations carried out on the basis of the dimerized spin Hamiltonian derived from first-principles.

\end{abstract}

\maketitle

\section{Introduction}
\label{sec:intro}

\begin{figure*} [t!] 
  \centering
  \includegraphics[width=1 \textwidth]{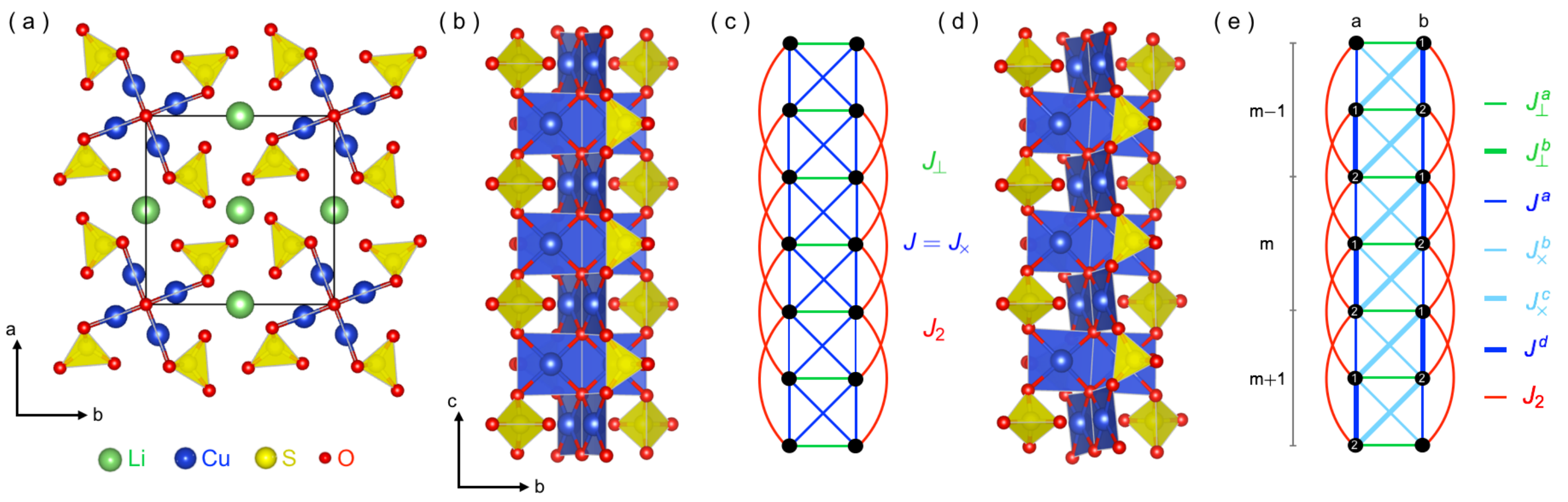}
  \caption{\label{fig:fig1} (a) Structure of tetragonal Li$_2$Cu$_2$O(SO$_4$)$_2$ viewed along the [001] 
  direction. (b) Structure of the [Cu$_2$O(SO$_4$)$_2$]$^{2-}$ chains running along the $c$ axis in the high-
  temperature ($T> 125 \text{ K} $) phase. (c) Schematic representation of the corresponding spin system. Intra 
  [Cu$_2$O$_6$]$^{8-}$ platelet rung coupling $J_{\perp}$ is represented in green, leg and diagonal couplings 
  $J = J_{\times}$ are represented in blue and next-nearest neighbor (NNN) couplings along the legs $J_2$ are in red. 
  (d) Structure of the [Cu$_2$O(SO$_4$)$_2$]$^{2-}$ chains in the distorted low-temperature ($T< 125 \text{ K}$) phase. 
  (e) Schematic representation of the corresponding staggered-dimer structure:  $J^a$ and $J^d$ in blue represent 
  the alternating couplings along the legs, $J_{\times}^b$ and $J_{\times}^c$ in light blue the diagonal inter-chain 
  couplings, $J_{\perp}^a$ and $J_{\perp}^b$ in green the couplings between the chains along the rungs and the 
  NNN coupling along the legs $J_2$ in red.}
\end{figure*}
Frustrated spin-1/2 two-leg ladders form a class of models which has been the subject of intense theoretical interest 
during the last decades as it fulfills all the requirements favoring the emergence of exotic
phenomena~\cite{lacroix2011introduction,Mila1998,Honecker2000,Vekua2006,Liu2008,RGHS2011,LiLin2012}. 
True material realizations of these models, however, are very scarce in the literature as the stringent conditions 
defining the geometry of frustrated ladders are rarely met in natural or synthetic compounds.  

A noticeable exception is BiCu$_2$PO$_6$, a system built from coupled spin-1/2 two-leg ladders where 
frustration arises from competing antiferromagnetic nearest and next-nearest neighbor interactions along 
the legs~\cite{Tsirlin2010}. The rich physics emerging from this particular geometry has triggered 
many experimental investigations of the magnetic properties of this compound over the past decade.  
These studies reveal a spin-gap behavior~\cite{Tsirlin2010,Koteswararao2007}, frustration-induced 
incommensurate dispersion of triplet quasiparticle excitations~\cite{Mentre2009,Plumb2013}, 
triplet dispersion renormalization resulting from a repulsion with the multi-quasiparticle continuum,~\cite{Plumb2016} 
or a cascade of field-induced phase transitions~\cite{Kohama2012} including, presumably, 
the formation and collapse of a quantum soliton lattice~\cite{Casola2013,Kohama2014}. 

A new material realization of a frustrated spin-1/2 two-leg ladder system, Li$_2$Cu$_2$O(SO$_4$)$_2$, 
has recently been discovered~\cite{Sun2015,Rousse2017,Vaccarelli2017}. At high temperature, this 
compound crystallizes in a tetragonal structure where [Cu$_2$O(SO$_4$)$_2$]$^{2-}$ 
chains running along the $c$ axis are well separated by Li$^+$ ions, thus forming quasi-1D structural 
units (see Fig.~\ref{fig:fig1}(a) and (b)). A first-principle investigation of the electronic structure of this 
compound~\cite{Vaccarelli2017} clearly identified these structural units as  magnetically equivalent 
to frustrated two-leg ladders where, as in BiCu$_2$PO$_6$, frustration arises from competing interactions 
along the legs. The geometry of this system, however, differs substantially from that of BiCu$_2$PO$_6$ 
because (i) inter-ladder interactions are very weak, (ii) the rung coupling ($J_{\perp}$ in Fig.~\ref{fig:fig1}(c)) 
is presumably ferromagnetic and (iii) additional antiferromagnetic diagonal interactions ($J_{\times}$ in 
Fig.~\ref{fig:fig1}(c)) occur. 

The phenomenology of this compound is also markedly different as a structural transition to a distorted triclinic 
phase occurs near 125~K~\cite{Rousse2017} (see Fig.~\ref{fig:fig1}(d)). Although lattice parameters are barely 
affected by this transition, large variations of the interplatelet Cu-O-Cu superexchange angles drastically 
impact the amplitudes of leg and diagonal interactions $J$  and $J_{\times}$. These couplings, equivalent 
by symmetry in the high-temperature phase, are largely split by the triclinic distortion, leading to the emergence 
of a dominant antiferromagnetic interaction $J^d$. The resulting structure, while maintaining the global geometry 
of a ladder, displays a strong magnetic dimerization, where dimers are staggered on the legs of the ladder, as 
shown in Fig.~\ref{fig:fig1}(e). This scenario, describing the evolution of a complex geometry with the temperature, 
is however essentially derived from first-principle calculations and a definite assessment of the model requires 
further experimental investigations.

In this work, we report the first detailed investigation of the low-temperature magnetic excitations of 
Li$_2$Cu$_2$O(SO$_4$)$_2$ combining inelastic neutron scattering (INS), magnetic susceptibility 
and infrared (IR) spectroscopy measurements carried out on powder samples. Dispersive triplet excitations 
are observed in INS above a spin gap $\Delta = 10.6 \text{ meV}$ at 1.5~K, a value consistent with the estimates 
extracted from magnetic susceptibility. Moreover, an absorption band visible only in the dimerized phase below 70~K 
is observed in IR spectroscopy and attributed to a triplet excitation arising at 14.3~meV. 
The dynamic Dzyaloshinskii-Moriya mechanism is invoked in this case to explain the 
absorption of light by this low-dimensional spin system. All these observations are qualitatively explained 
by higher-order perturbation and exact diagonalization calculations of triplet quasi-particle excitations carried 
out on the basis of the dimerized geometry derived from previous first-principle calculations~\cite{Vaccarelli2017}. 

\section{Experiments} \label{sec:exp}

\subsection{Time-of-flight neutron spectroscopy}

Inelastic neutron scattering experiments were performed at the Institut Laue-Langevin in Grenoble 
on the thermal time-of-flight spectrometer IN4 using incoming energies of 16.6, 32.0, and 66.6 meV. 
Most measurements were done with an incoming energy of 32.0 meV from the (004) reflection of a 
pyrolytic graphite monochromator. A Fermi chopper was used with a $2^\circ$ straight slit package 
spinning at 15000~rpm. The energy resolution (full width at half maximum) at elastic energy transfer 
was 2.1 meV. A total of $6.8 \text{ g}$ of powder sample of Li$_2$Cu$_2$O(SO$_4$)$_2$
was synthesized according to the procedure described in Ref.~\onlinecite{Sun2015}, put in a flat aluminum 
plate holder of dimensions of $40 \times 28 \times 2 \text{ mm}^{3}$, and mounted in a standard 
helium cryostat. Measurements were carried out at temperatures $T = 1.5$,  $40$, $60$, $80$ and 
$100 \text{ K}$ for a typical acquisition time of $12 \text{ h}$ per temperature.
The scattering from the empty Al-holder was measured and subtracted from the data, 
which was subsequently corrected for self-shielding and absorption using standard data reduction routines.
Data treatments were carried out using the LAMP software~\cite{LAMP1996}.

\begin{figure}[h]
  \includegraphics[width=0.48 \textwidth]{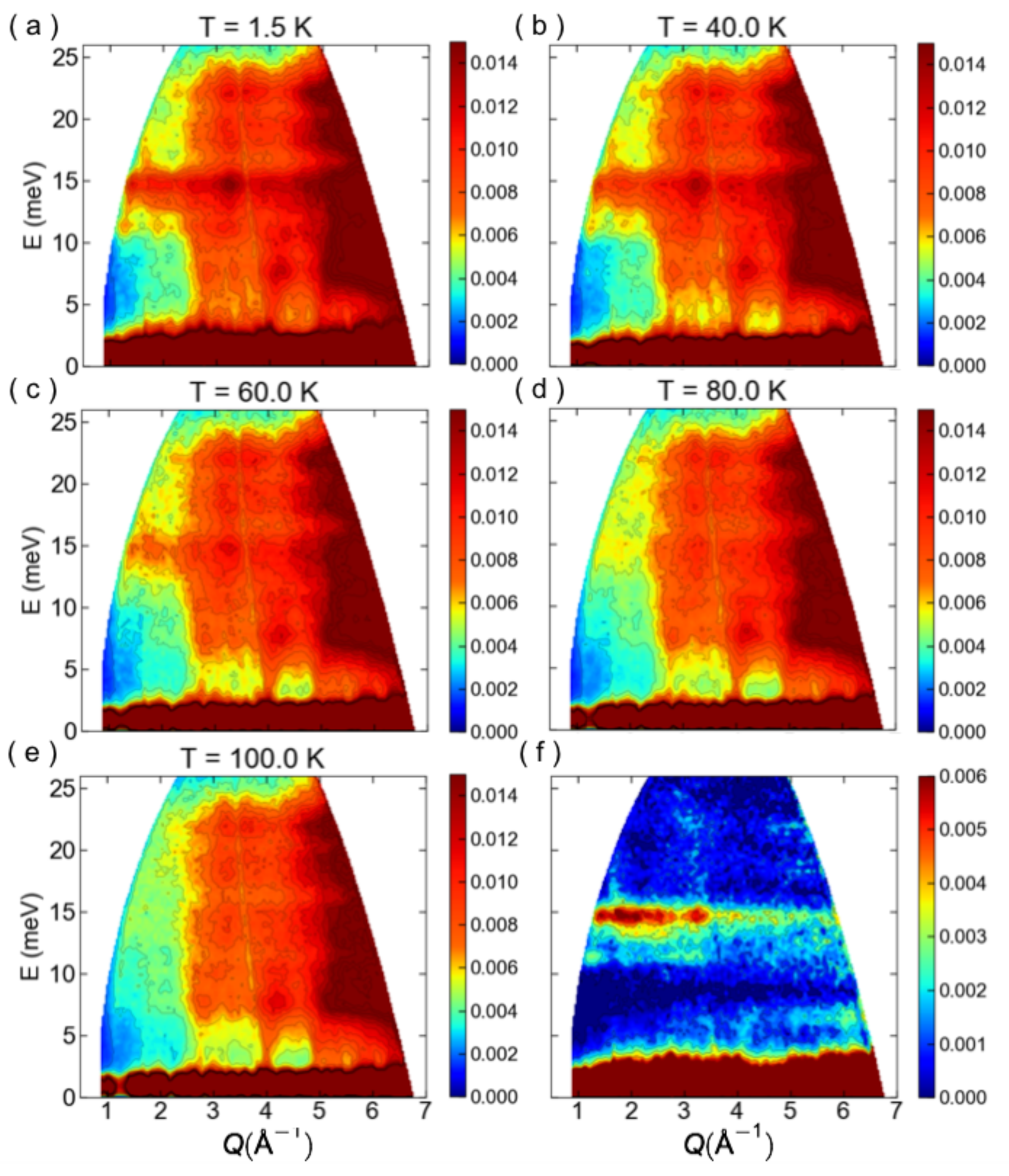}
  \caption{\label{fig:fig2} Experimental dynamic susceptibility $\chi''(|Q|,E)$ plots for Li$_2$Cu$_2$O(SO$_4$)$_2$ 
  measured at different temperatures: (a) $T=1.5$~K, (b) $T=40$~K, (c) $T=60$~K, (d) $T=80$~K, and (e) $T=100$~K.
  Magnetic contributions are isolated in (f) by displaying the difference $\Delta \chi''(|Q|,E)$ from Eq.~(\ref{eq:deltachi}).} 
\end{figure}
Figure \ref{fig:fig2}(a)-(e) shows the maps of the dynamic susceptibility $\chi''(|Q|,E)$, obtained 
by normalizing the background-subtracted neutron scattering intensity $S(|Q|,E)$ by the thermal occupancy 
factor $1-e^{-E/k_{\mathrm{B}}T}$ at 1.5, 40, 60, 80, and 100~K, respectively. The temperature dependence of the dynamic 
susceptibility clearly reveals the presence of two dominant contributions arising from phonon and magnetic 
excitations. Whereas the scattering cross-section of the former scales as $|Q|^2$, 
that arising from magnetism scales with the square of the form factor associated with the magnetic ions, 
and falls off with increasing $|Q|$ ~\cite{Lovesey1984,Squires1978}. The high-temperature dynamic susceptibility is therefore 
largely dominated by phonon scattering (see Fig.~\ref{fig:fig2}(d)-(e)) whereas the weight of the magnetic 
contribution progressively increases with decreasing temperatures, as it can be observed in the low-$|Q|$ region 
of Fig.~\ref{fig:fig2}(a)-(c). Assuming, in first approximation, that the intensity is entirely associated with phonons 
scattering for the highest temperature measurement ($T=100$~K in our case), the magnetic contribution to the 
dynamic susceptibility at low-temperature can simply be isolated by plotting the difference~\cite{Clancy2011}
\begin{equation}
\label{eq:deltachi}
\Delta \chi''(|Q|,E) =  \chi''(|Q|,E)|_{1.5\mathrm{K}} - \chi''(|Q|,E)|_{100\mathrm{K}}.
\end{equation}
\begin{figure}[h]
  \centering
  \includegraphics[width=0.48 \textwidth]{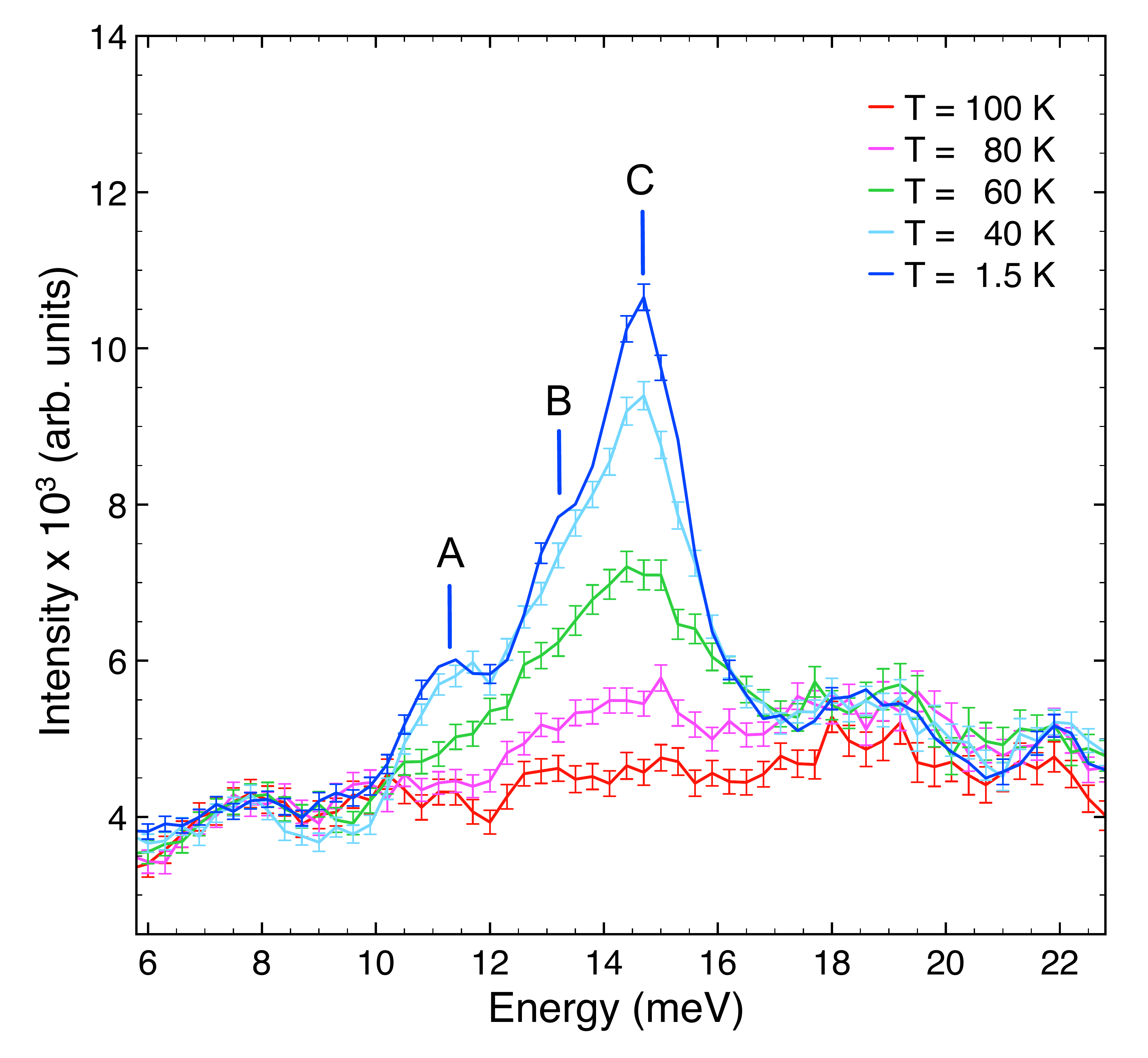}
  \caption{ \label{fig:fig3} Constant-$Q$ cuts of the dynamic susceptibility  $\chi''(|Q|,E)$ extracted for a $|Q|$ 
  value of $2 \pm 0.1 \text{ \AA}^{-1}$ from the plots shown in Fig.~\ref{fig:fig2}(a)-(e).}
\end{figure}
The corresponding difference map is shown in Fig.~\ref{fig:fig2}(f) and clearly reveals the presence of  
dispersive magnetic excitations of bandwidth $\approx\text{5 meV}$ above a large spin gap $\Delta = 10.6 \text{ meV}$, 
measured as the inflection point of the first peak (A). 
Constant-$Q$ cuts extracted for a $|Q|$ value of $2 \pm 0.1 \text{ \AA}^{-1}$ for each temperature measured experimentally 
are shown in Fig.~\ref{fig:fig3}. The broad band of magnetic excitations, extending over 5~meV above the spin gap, includes 
the first low-intensity peak (A) centered at $11.2 \text{ meV}$ followed by the most intense structure (C) at $14.7 \text{ meV}$. 
A low-energy shoulder (B) of this dominant structure is also present at about $13.5 \text{ meV}$, particularly visible 
on the data obtained at 1.5~K. Inelastic neutron scattering experiments therefore clearly reveal the presence 
of dispersive magnetic excitations, mostly localized at low-$|Q| < 3.5 \text{ \AA}^{-1}$ and vanishing at temperatures 
above $\sim 80 \text{ K}$, consistent with spin triplet excitations out of the reported singlet ground state of this 
compound~\cite{Rousse2017,Vaccarelli2017}.

\subsection{Static magnetic susceptibility} 

\begin{figure} [t!]
 \centering
 \includegraphics[width=0.48 \textwidth]{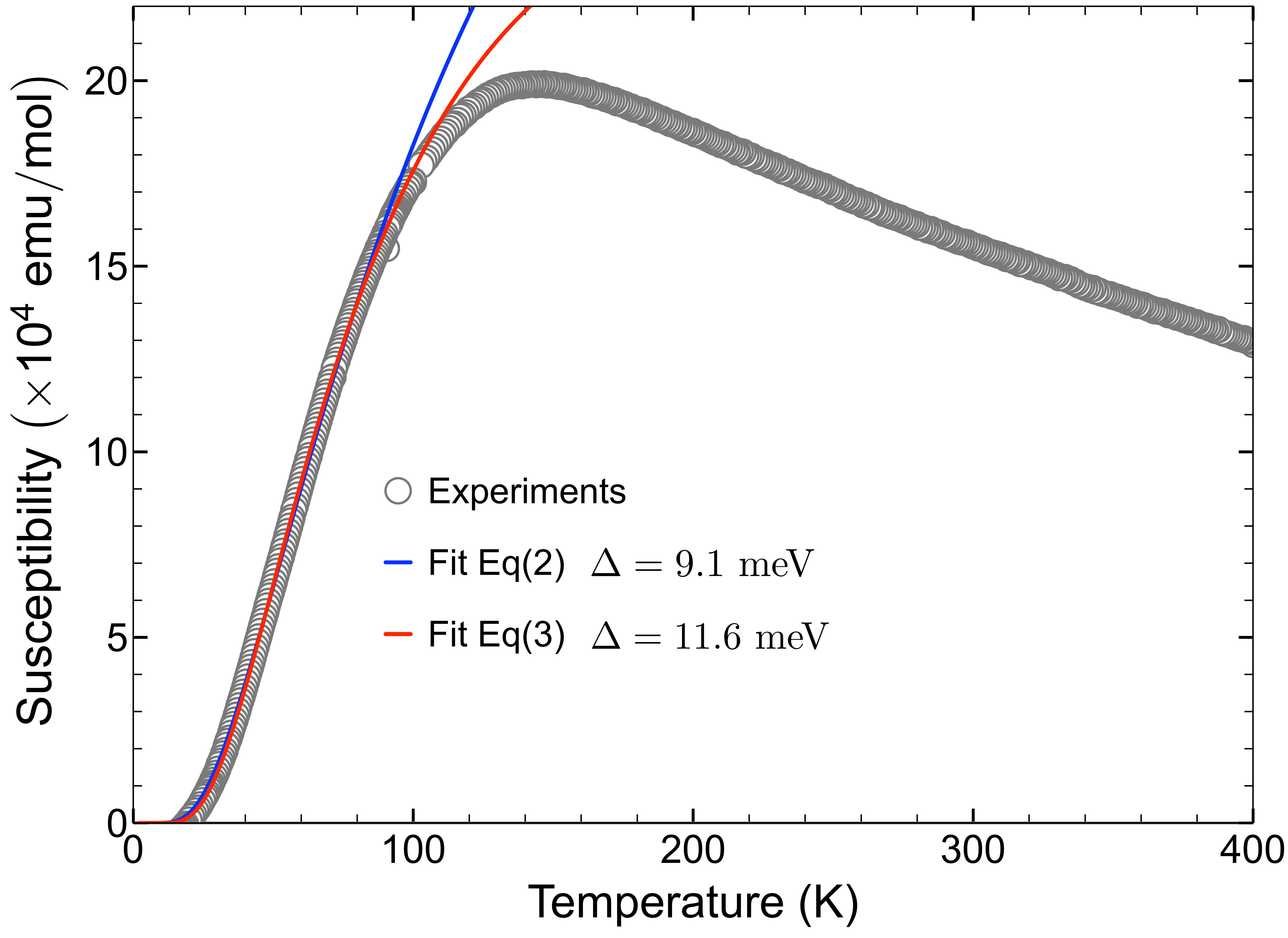}
 \caption{\label{fig:fig4} Temperature dependence of the static magnetic susceptibility of  
 Li$_2$Cu$_2$O(SO$_4$)$_2$ corrected from paramagnetic impurities and 
 temperature-independent contributions. Grey circles correspond to the experimental 
 points, the solid blue and red lines correspond, respectively, to the best fit obtained 
 using the general expression~(\ref{equ:general}) or a spin-1/2 two-leg ladder Heisenberg 
 model~(\ref{equ:ladder}). The corresponding values of the extracted spin gap are indicated.}
\end{figure}

These results can be further confirmed by analyzing the low-temperature behavior of the experimental 
static magnetic susceptibility. Fig.~\ref{fig:fig4} reproduces the experimental data corrected from paramagnetic 
impurities and temperature-independent contributions, already reported in 
Ref.~\onlinecite{Rousse2017}. The susceptibility exhibits the typical behavior of a 
gapped low-dimensional antiferromagnet with a high-temperature paramagnetic regime reaching a broad 
maximum at about 125~K and an exponential decay at low-temperature. Note, however, that the 
temperature dependence of the magnetic susceptibility is largely affected by the structural transition 
occurring at the same temperature as the maximum ($125 \text{ K}$) since it is accompanied by a substantial 
magnetic dimerization within the ladders. As reported in Ref.~\onlinecite{Vaccarelli2017}, although 
this transition extends over a large temperature range, the low-temperature triclinic phase is already 
mostly stabilized at about 80~K, and the 2-80~K range can therefore be used to estimate the 
corresponding spin gap. 
A rough estimate can be obtained by fitting the experimental data using the general expression for thermally 
activated processes
\begin{equation} \label{equ:general}
\chi(T) \propto  e^{-\Delta/T}
\end{equation}
leading to a value of $\Delta \approx 9.1 \text{ meV}$. It should however be noted that
(\ref{equ:general}) would be valid for non-interacting particles, but
magnons are not free particles.
A suitable expression that takes the hard-core repulsion between the
magnons into account is given by~\cite{Troyer1994}
\begin{equation} \label{equ:ladder}
\chi_{\mathrm{lad.}}(T) \propto  T^{-1/2} \; e^{-\Delta/T}
\end{equation}
for a quadratic band minimum. Equation (\ref{equ:ladder}) has 
also been employed (see Fig.~\ref{fig:fig4}). The resulting estimate for the spin gap, 
$\Delta \approx 11.6 \text{ meV}$, is slightly larger than the value obtained with the previous expression. 
Despite these small variations, essentially related to the rather low accuracy of this approach and to the limited 
applicability of the simple ladder model to Li$_2$Cu$_2$O(SO$_4$)$_2$, these estimates are however 
fully consistent with the spin gap value obtained from inelastic neutron scattering.

\subsection{IR spectroscopy} \label{subsec:IR}

Infrared absorption spectroscopy was finally employed as a complementary technique to investigate the 
low-energy excitations of Li$_2$Cu$_2$O(SO$_4$)$_2$. These experiments were carried out using a Fourier 
transform Bruker IFS66 v/s spectrometer at the IMPMC-Sorbonne University spectroscopy platform. The instrument was  
aligned in transmission geometry. Isotropic pellets of $\sim 13 \text{ mm}$ 
diameter were prepared by mixing the original powder samples with transparent matrix materials. Pure KBr powder 
was used as a matrix for the pellets employed in the middle-infrared (MIR), whereas polyethylene (PE) 
was employed to prepare pellets for the far infrared (FIR) measurements. The former were obtained by mixing $2.5 \text{ mg}$ 
of sample and $80 \text{ mg}$ of pure PE; for the latter, $1.4 \text{ mg}$ of sample was mixed with $200 \text{ mg}$ 
of KBr powder, placed in an oven at $T = 150^{\circ} \text{C}$ in order to remove water contamination from the 
KBr powder, and then pressed to obtain high quality pellets. Transmission spectra were taken as a function of temperature from $10$ to 
$300 \text{ K}$ using a continuous Janis liquid helium cryostat working in vacuum. 
Each spectrum was acquired in the frequency region $60-640 \text{ cm}^{-1}$ for the FIR measurements 
and $580 - 4400 \text{ cm}^{-1}$ for the spectrum in the MIR, with a spectral resolution of about $2 \text{ cm}^{-1}$. 
The FIR spectra were recorded with a DTGS-Pe detector and a multilayer mylar beamsplitter. The MIR data were obtained 
with a HgCdTe (MCT) detector and a Ge coated KBr beamsplitter. For both region we used a Globar (SiC) source. 

\begin{figure} [t!] 
  \centering
  \includegraphics[width=0.48 \textwidth]{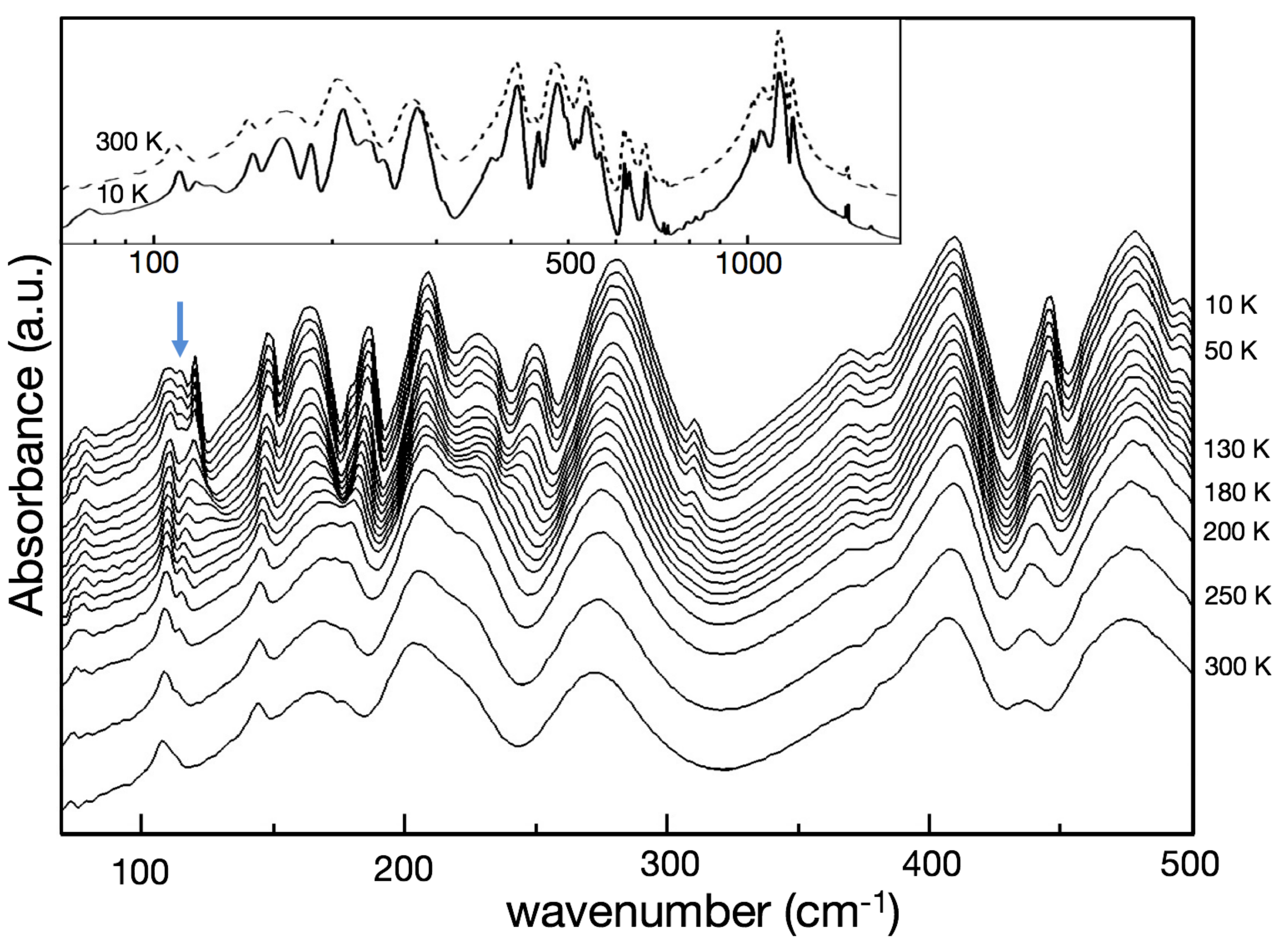}
  \caption{\label{fig:fig5} Temperature dependence of the powder transmission infrared spectra of 
  Li$_2$Cu$_2$O(SO$_4$)$_2$ in the FIR region. The blue arrow indicates the spectral feature associated to 
  the magnetic excitation. Top inset: The lowest and highest temperature spectra are reported in logaritmic 
  scale up to $1800 \text{ cm}^{-1}$ to show the high energy modes associated with the internal [SO$_4$]$^{2-}$ bond stretching.}
\end{figure}

The temperature dependence of the transmission IR powder spectrum of  Li$_2$Cu$_2$O(SO$_4$)$_2$  
measured between $10$ and $300 \text{ K}$ is shown in Fig.~\ref{fig:fig5}. Most of the absorption bands visible 
between $100$ and $1300 \text{ cm}^{-1}$ are associated with the electric dipole excitation of optical phonons. 
The group of high-energy modes located around $1100 \text{ cm}^{-1}$ is exclusively associated with internal 
[SO$_4$]$^{2-}$ bond stretching~\cite{Farmer1974}. The $500-700 \text{ cm}^{-1}$ range is dominated by 
[SO$_4$]$^{2-}$ tetrahedra bending modes involving progressively the displacement of Cu and O atoms 
forming the chain backbone, as the frequency decreases.

Symmetry can be employed to determine the maximum number of infrared active optical phonons in both phases
of this compound. In its high temperature phase, Li$_2$Cu$_2$O(SO$_4$)$_2$ belongs to the $P4_2/m$ space group.  
A factor group analysis indicates that the vibrational degrees of freedom decompose as 
\begin{equation}
\Gamma_{\text{HT}} = 11 A_g + 13 B_g + 9 E_g + 10 A_u+ 8 B_u + 15 E_u
\end{equation}
on the irreducible representations of the $D_{4h}$ point group. Subtracting the $ A_u + E_u$ acoustic modes,
we find $37$ IR-active modes, decomposed in $23$ potentially distinct $9 A_u + 14 E_u$ bands. The triclinic distortion, 
occurring below $125 \text{ K}$, further reduces the crystal symmetry to  $P\bar{1}$. A similar analysis leads to the 
following decomposition on the only two irreducible representations of $C_i$
\begin{equation}
\Gamma_{\text{LT}} = 42 A_g + 48 A_u
\end{equation}
leading, after subtraction of the $3 A_u$ acoustic modes, to $45 A_u$ IR-active modes.
As it can be observed in Fig.~\ref{fig:fig5}, the exact number of bands detected in these experiments is difficult to assess,  
due to the broad and asymmetric profile of certain peaks. The above group theoretical analysis therefore only provides an upper 
bound for the number of bands distinguishable in the experimental spectra. Qualitatively, however, the large increase 
of active modes due to the symmetry lowering triggered by the triclinic distortion is clearly visible on the experimental spectra 
when decreasing the temperature below the transition ($\sim 125 \text{ K}$) and therefore consistent with the structural 
data~\cite{Rousse2017}. 

\begin{figure} [t!]
  \centering
  \includegraphics[width=0.48 \textwidth]{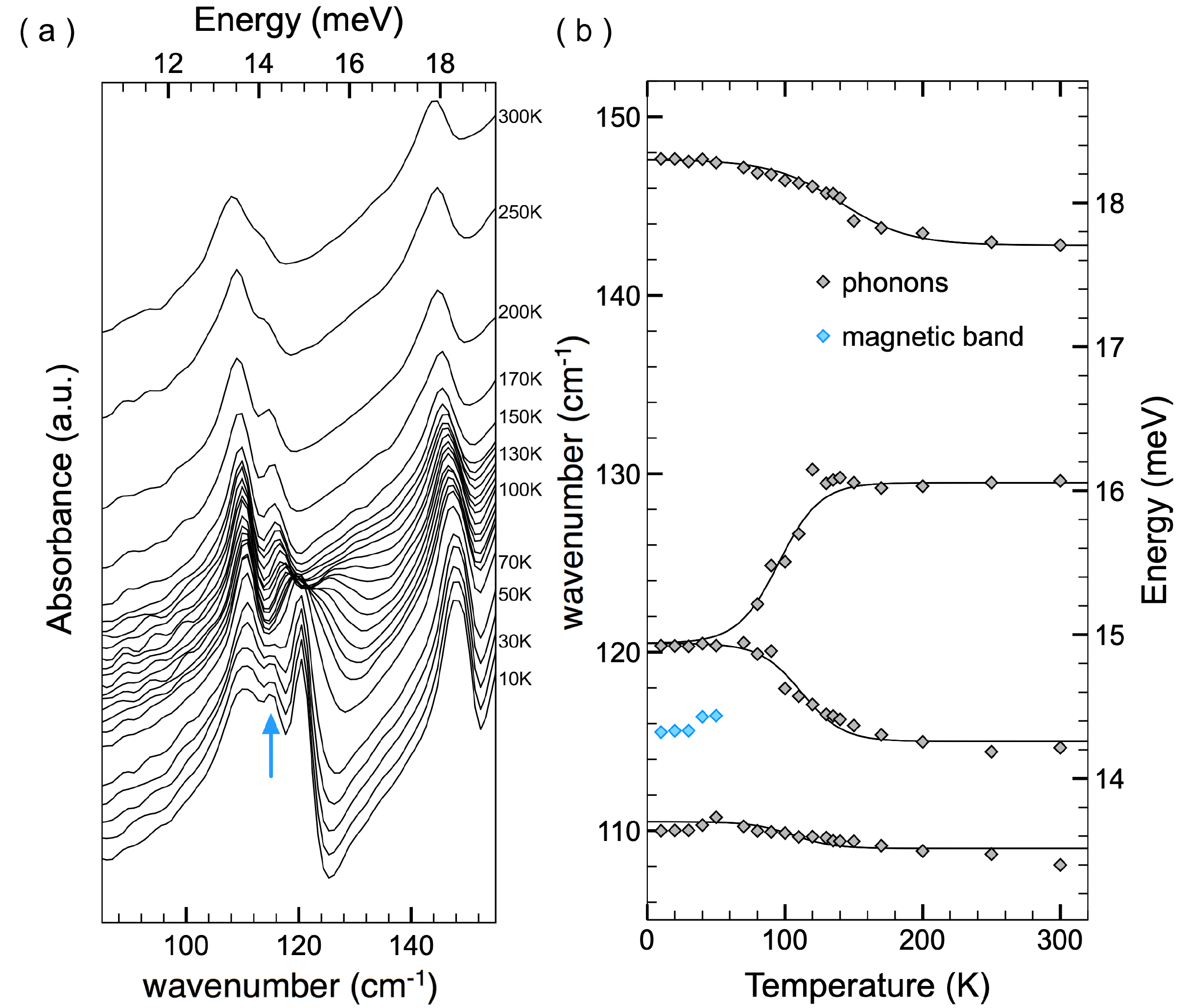}
  \caption{\label{fig:fig6} (a) Enlarged view of the powder transmission infrared spectra in the $80 - 155 \text{ cm}^{-1}$ range. 
(b) Temperature-dependence of the mode frequencies extracted from the Lorentzian fit. Experimental data are represented with filled 
diamond symbols, lines are only guides for the eye. The phonon bands are represented in black whereas the magnetic band is in blue.} 
\end{figure}

Besides this increase in the number of phonon lines, the main effects of decreasing the temperature consist in a slight 
hardening and narrowing of most of the bands, usually attributed with anharmonic effects and in particular, for the former, 
with the overall unit cell volume contraction~\cite{Rousse2017}. However, a few bands display a softening in the temperature 
range of the structural transition, characteristic of magneto-elastic effects associated in our case to the rise of the 
dimerization \cite{Room2004b}. At low energy, however, a pronounced transfer of spectral weight toward low frequencies 
reveals the rise of a weak excitation at $115 \text{ cm}^{-1}$ ($14.3 \text{ meV}$). This band is indicated by a blue 
arrow in Fig.~\ref{fig:fig6}(a). 

In order to quantify these spectral changes, a least-square fit of the low-frequency range of the spectra based on a 
superposition of Lorentzians has been carried out. The temperature dependence of the energy of the different modes 
observed in the $105 - 152 \text{ cm}^{-1}$ range resulting from this fit is displayed in Fig.~\ref{fig:fig6}(b). 
A weak hardening of the modes identified as polar phonons (grey filled diamonds) is visible with decreasing temperature 
except in the transition temperature range ($80-130 \text{ K}$) where a sizable jump is observed. Concomitantly, the band 
shown in blue in Fig.~\ref{fig:fig6}(b) has an energy of $14.3 \text{ meV}$, that falls precisely in the broad band 
of magnetic excitations observed by INS, in a region characterized by a large spectral weight. Moreover, this excitation is 
only visible at temperatures well below the structural transition, \textit{i.e.} in the magnetic dimerized phase. These 
observations therefore suggest that this excitation might involve, to a certain extent, the spin degrees of freedom of this system.

\section{Theory} \label{sec:theo}

In order to explain this set of experimental findings and explore the landscape of low-energy magnetic excitations in 
Li$_2$Cu$_2$O(SO$_4$)$_2$, both exact diagonalization and higher-order perturbation theory calculations have been carried out. 
The $Q$-dependence of the dynamic susceptibility $\chi''$ should be dominated by that of the $J^d$ dimers~\cite{Furrer1979}, 
or more generally be related to the exchange constants via a first-moment sum rule~\cite{Hohenberg1974,Stone2002,Quintero-Castro2010,Canevet2015}. 
However, any such analysis is complicated in Li$_2$Cu$_2$O(SO$_4$)$_2$ by the large phonon background. We have 
therefore decided to focus rather on the clear magnetic features A, B, and C visible in Fig.~\ref{fig:fig3} by comparing to both 
exact diagonalization and higher-order perturbation theory calculations. 

A realistic spin Hamiltonian (see Appendix~\ref{appA}) capable of describing the magnetism of this compound 
in the triclinic phase has been derived previously~\cite{Vaccarelli2017} from first-principle calculations and requires seven distinct 
couplings to account for the low symmetry of the crystal. The geometry of this model is depicted in Fig.~\ref{fig:fig1}(e) where 
$J^d = 1$ and  $J^a = 20/330 = 2/33$ are antiferromagnetic and alternate along the legs of the ladder so as to form a staggered 
dimer structure,   $J_{\perp}^a \approx J_{\perp}^b = -110/330 = -1/3$ are the ferromagnetic couplings along the rungs of the ladder, 
$J_{\times}^b = 78/330 = 13/55$ and $J_{\times}^c = 133/330$ are antiferromagnetic diagonal couplings between the legs and, finally, 
$J_2 = 112/330 = 56/165$ is the antiferromagnetic NNN interaction along the legs. This model therefore neglects the supposedly 
very weak inter-ladder couplings~\cite{Vaccarelli2017} as well as any other term beyond the bilinear, Heisenberg like, interactions. 

Exact diagonalization (ED) calculations have been performed using finite lattices of $N=12$, 16, 20, 24, 28, and 32 sites with 
periodic boundary conditions along the legs, as the one-dimensional magnetic unit cell contains two dimers, \textit{i.e.} four spins.
For system sizes exceeding $N=20$ we have used the Lanczos algorithm~\cite{Lanczos1950,Dagotto1994} in order to compute 
low-lying eigenvalues. Furthermore, perturbation expansion of the one-triplet dispersion relation up to the fifth order has 
been carried out around the limit of isolated dimers.
In this approach, the unperturbed ground state corresponds to a product of singlets, 
$ \left| s \right. \rangle = ( \left| \uparrow \downarrow \right. \rangle - \left| \downarrow \uparrow \right. \rangle ) / \sqrt{2} $,
on the leg dimers defined by the dominant antiferromagnetic coupling $J^d$.
Low-energy magnetic excitations of this system are obtained by promoting one dimer into a triplet state,  
$ \left| t_{-1} \right. \rangle = \left| \downarrow \downarrow \right. \rangle$,  
$ \left| t_0 \right. \rangle = ( \left| \uparrow \downarrow \right. \rangle +  \left| \downarrow \uparrow \right. \rangle ) / \sqrt{2} $ or 
$ \left| t_{1} \right. \rangle = \left| \uparrow \uparrow \right. \rangle$. 
Since the unit cell contains two dimers, we find two separate single-particle bands, as shown in Fig.~\ref{fig:fig7} 
(see Appendix~\ref{appA} for details).

\begin{figure*} [t!]
  \centering
  \includegraphics[width=1 \textwidth]{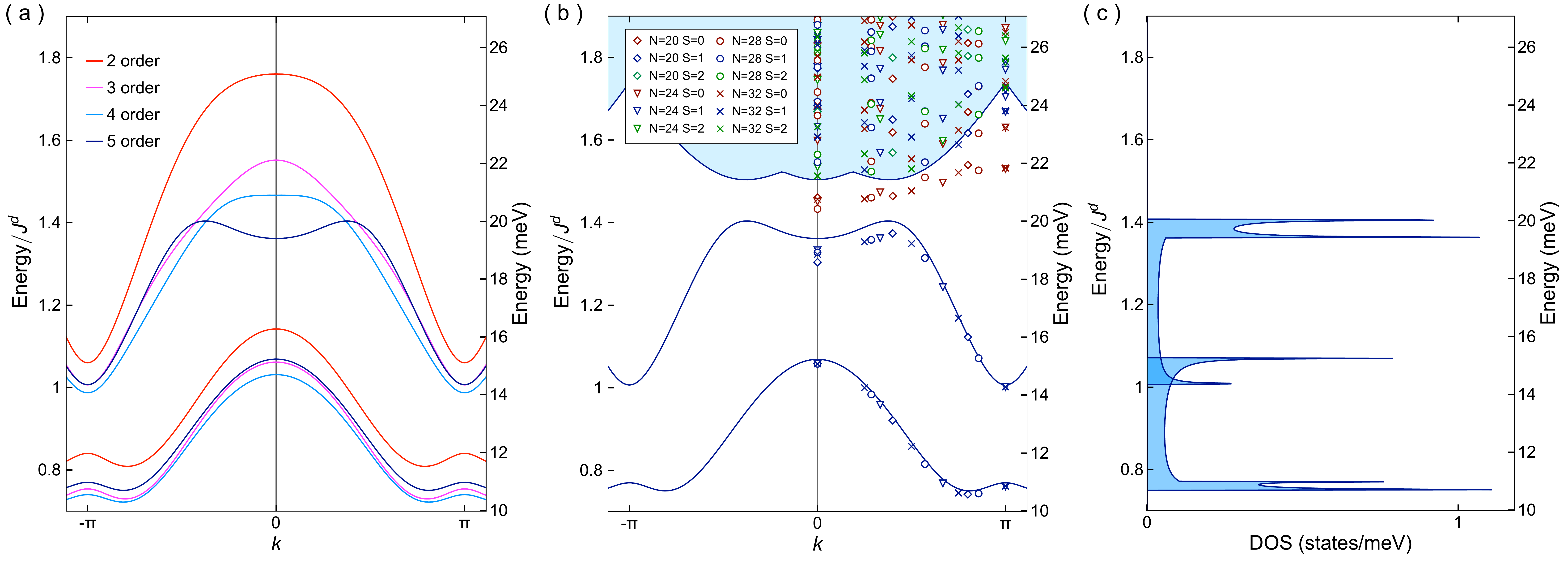}
  \caption{\label{fig:fig7}(a) One-triplet dispersion relation calculated from second to fifth order perturbation theory. (b) Comparison 
  of the one-triplet dispersion relation obtained for the fifth order perturbation theory with exact diagonalization results obtained on finite 
  lattices up to $N=32$
sites. First few lowest-lying singlets (in red) and triplet (in blue) obtained from ED are shown. The blue filled area
  corresponds to the free two-triplet continuum. (c) One-triplet density of states obtained from fifth order perturbation theory dispersion 
  relations shown in (b). Energies given in $\text{meV}$ have been obtained by downscaling the DFT isotropic magnetic couplings by 50~\%.  }
\end{figure*}

Numerical results are summarized in Fig.~\ref{fig:fig7}. Fig.~\ref{fig:fig7}(a) and (b) show the convergence of the perturbation 
expansion by comparing, respectively, the one-triplet dispersion relations obtained at different expansion orders and the 
highest order perturbation theory with exact diagonalization results. As it clearly appears, a remarkable convergence 
towards the exact results is achieved for the higher-order expansions. These calculations indicate the presence of two 
dispersive and slightly overlapping triplet bands above a large spin gap. The lowest band displays a behavior characteristic  
of antiferromagnetically coupled dimers with a maximum at the zone center and a minimum close the Brillouin zone edge. 
The actual minimum arises at an incommensurate wave vector and results from the presence of frustrating couplings.
In addition to the one-triplet excitation bands, the lower boundary of the two-triplet continuum is calculated as
\begin{equation}
E_2(k) = \min_{\substack{q \in \mathrm{1BZ} \\ m,n = 1,2}} \left[  \omega_m(k-q) +w_n(q) \right]
\end{equation}
where $\omega_1(k)$ and $\omega_2(k)$ represent the two one-triplet bands, is also shown in Fig.~\ref{fig:fig7}(b). The 
large value of the spin gap compared to the modest triplet excitation bandwidth, pushes this continuum lower bound well above the 
maximum of the highest one-triplet band. The ED results are close to the fifth-order expansion, \textit{i.e.}, both of them can be 
considered accurate. The exception is the top of the upper band where proximity to the continuum leads to larger finite-size 
effects and slower convergence of the series (see Appendix~\ref{appB} for details). Exact diagonalization, furthermore, reveals 
the presence of lower-lying singlets above and below the continuum, which do not interfere with the upper triplet band, as it can 
be seen in Fig.~\ref{fig:fig7}(b). Similar excitations have already been reported in ladder systems where they
can be understood as bound states of two triplets~\cite{Trebst2000,Windt2001,Tsirlin2010}.

\section{Discussion} \label{sec:disc}

The theoretical results presented in the previous section provide solid ground for analyzing the experimental results obtained 
on Li$_2$Cu$_2$O(SO$_4$)$_2$. It should be noted, however, that the global energy scale obtained from first-principle calculations~\cite{Vaccarelli2017} 
is not consistent with our experimental observations. Indeed, as already reported, 
a straightforward use of the magnetic couplings provided by density functional theory (DFT) calculations
leads to a substantial overestimation of the experimental spin gap~\cite{Vaccarelli2017}. Although the amplitudes of these couplings
are often overestimated and strongly depend on the approximate treatment of exchange and correlation employed in the 
calculations~\cite{Martin1997,Saul2014}, their ratios are expected to be subject to smaller errors~\cite{Jeschke2011}. In this 
framework, the ratios between the seven couplings involved in the spin Hamiltonian were considered as fixed. The global energy 
scale was thus taken as the only variable parameter, adjusted so as to reproduce the experimental value of the spin gap. This led to an 
approximate 50~\% downscaling of the DFT coupling amplitudes. The resulting energy scale in millivolts is shown on the vertical
axes of Fig~\ref{fig:fig7}(a) to (c).

Under these assumptions, powder INS can be qualitatively discussed in terms of one-triplet excitation density-of-states (DOS) shown in 
Fig.~\ref{fig:fig7}(c). In first approximation, the experimental peaks A, B and C shown in Fig.~\ref{fig:fig3} can indeed be interpreted 
as arising from the DOS singularities, at the bottom of the lowest band for peak A and in the overlapping region of the two bands 
for peaks B (bottom of the upper band) and C (top of the lower band). Although this qualitative analysis provides a satisfying explanation
regarding the origins of the low-energy part of the INS data, it also predicts the presence of higher-lying features corresponding to the top
of the highest one-triplet band, \textit{i.e.}\ at $\sim 20 \text{ meV}$, which was not observed experimentally. This could simply be a matrix 
element effect. Alternatively, although our model locates the two-triplet continuum lower bound above the highest 
one-triplet branch over the entire first Brillouin zone (see Fig.~\ref{fig:fig7}(b)), they remain close in energy. Therefore, only minor 
modifications of the model employed in this work would be necessary to change this picture and, in particular, restore a significant overlap 
between the highest triplet quasi-particle mode and the two-particle continuum. This overlap will provide
spontaneous decay channels~\cite{Zhitomirsky2006,Zhitomirsky2013} leading to significant damping of these quasi-particles and therefore 
to the absence of visible signatures in INS data. 

A second important question arises from the likely detection of triplet excitations in IR spectroscopy presented in Sec.~\ref{subsec:IR}. 
Indeed, dominant electric dipole transitions induced by light are strictly confined to spin-conserving excitations ($\Delta S = 0$)
and are therefore, in principle, unable to reveal singlet-to-triplet transitions. However, it has been shown that, in a number of  low-dimensional 
quantum magnets, this selection rule can be circumvented through essentially two mechanisms relying on the presence 
of spin-phonon coupling and involving one or multiple magnetic excitations. 

A successful and now well-established model employed to describe the infrared optical absorption of one and two-dimensional undoped
cuprates is based on phonon-assisted bi-magnon absorption~\cite{Lorenzana1995a,Lorenzana1995b,Lorenzana1997,Suzuura1996,Windt2001}. 
The excitation of singlet bound states, resulting from the coupling of two spin-carrying modes (triplets, in our case) 
in such a way that the total spin amounts to zero, indeed obeys the imposed spin selection rule. Lorenzana and Sawatzky
further showed that, when a center of inversion is present, dipole-allowed absorption is only possible if a symmetry-breaking 
phonon is also involved in the process~\cite{Lorenzana1995a,Lorenzana1995b}. In our case, an attribution of the 
IR band observed at $14.3 \text{ meV}$ for $T <70 \text{ K}$ to the absorption of phonon-assisted bi-magnons is very unlikely as the typical 
energy of these excitations, already of the order of $\sim 2 \Delta = 21.2 \text{ meV}$ when neglecting the phonon energy, are much larger. 
 
An alternative mechanism, arising from the spin-orbit coupling, has been proposed to explain the detection of singlet-to-triplet 
excitations in dimerized quantum magnets using IR absorption~\cite{Room2004a,Room2004b}. It can be described qualitatively 
as a process where light excites the system into a virtual spin-singlet one-phonon state coupled, through a \textit{dynamic} 
Dzyaloshinskii-Moriya (DM) interaction, to a spin-triplet zero-phonon state~\cite{Cepas2004}. This mechanism thus relies on 
the assumption that the virtual polar phonon involved in the process is associated with atomic displacements able to induce an 
instantaneous variation of the DM vector.  Assuming that such a mechanism is effective in the low-temperature phase 
of Li$_2$Cu$_2$O(SO$_4$)$_2$, the $14.3 \text{ meV}$ IR absorption band would, quite accurately, match the zone 
center maximum of the lowest one-triplet excitation and the corresponding Van Hove singularity in the DOS.

\section{Conclusion}

In conclusion, we report the first experimental investigation of magnetic excitations in the low-temperature, dimerized phase 
of the recently discovered frustrated spin-1/2 two-leg ladder Li$_2$Cu$_2$O(SO$_4$)$_2$.  Through a combined analysis 
of inelastic neutron scattering, magnetic susceptibility and infrared absorption spectroscopy data obtained on powder samples, 
dispersive  excitations of bandwidth of the order of $5 \text{ meV}$ have been clearly identified above a large spin gap 
of $10.6 \text{ meV}$. Exact diagonalization and higher-order perturbation theory calculations allowed for an overall 
consistent interpretation of these results in terms of one-triplet quasi-particle excitations above the singlet ground state. 
While experiments and theory show an overall good agreement, the only exception lies in the high-energy part of the triplet excitation
spectrum, where a possible coupling between the quasi-particles and the high-lying many-particle continuum may be responsible 
for the absence of high-energy structure in the INS spectra. This calls for further experimental and theoretical investigations 
of this very rare example of frustrated spin-1/2 ladder, which will heavily rely on the future availability of single crystals.

\begin{acknowledgments}

This work was supported by French state funds managed by the ANR within the Investissements d'Avenir programme under
reference ANR-11-IDEX-0004-02, and more specifically within the framework of the Cluster of Excellence MATISSE led by 
Sorbonne Universit\'es. The authors thank Dr Meiling Sun for providing the sample used in this study. Inelastic 
neutron scattering experiments were performed at the Institut Laue-Langevin (ILL) in Grenoble. 

\end{acknowledgments}

\appendix

\section{Series expansions}\label{appA}

The evaluation of the one-triplet dispersion relation has been carried out solving the spin Hamiltonian parametrized 
from first-principles around the limit of isolated dimer (low-temperature phase), implementing a high-order perturbative 
approach in the strong coupling expansion in the spirit of textbooks such as chapter 11 of Ref.~\onlinecite{Baym}.

Following Ref.~\onlinecite{Vaccarelli2017}, the triclinic phase can be described by the staggered $S=1/2$ dimer structure schematized 
in Fig.~\ref{fig:fig1}(e). In this structure one of the couplings along the legs, $J^d$, is much stronger than all the others.
The triclinic Hamiltonian $H$ can thus be written as the sum of an unperturbed part ${H}_0$, for decoupled dimers along the legs
(bold blue lines in Fig.~\ref{fig:fig1}(e)) and a perturbation, ${W}$, accounting for the coupling between the dimers with
\begin{equation}
{H}_0 = \sum_m \bigg[ J^d \Big( \textbf{S}_{1,a}^m \cdot \textbf{S}_{2,a}^{m} + \textbf{S}_{1,b}^{m} \cdot \textbf{S}_{2,b}^{m} \Big) 
\bigg]
\end{equation}
and 
\begin{widetext}
\begin{equation}\label{eq:PT_per_tri}
\begin{split}
{W} = & \sum_m  \bigg[ J_{\perp}^a \Big( \textbf{S}_{1,a}^m \cdot \textbf{S}_{2,b}^m \Big) + J_{\perp}^b \Big( \textbf{S}_{2,a}^{m-1} 
\cdot \textbf{S}_{1,b}^m \Big) +J_{2} \Big( \textbf{S}_{1,a}^m \cdot \textbf{S}_{1,a}^{m+1} + \textbf{S}_{2,a}^m \cdot \textbf{S}_{2,a}^{m+1}  
+ \textbf{S}_{1,b}^m \cdot \textbf{S}_{1,b}^{m+1} + \textbf{S}_{2,b}^m \cdot \textbf{S}_{2,b}^{m+1} \Big)  \\
&+J^a \Big( \textbf{S}_{1,a}^m \cdot \textbf{S}_{2,a}^{m-1} + \textbf{S}_{1,b}^{m} \cdot \textbf{S}_{2,b}^{m-1} \Big) +J_{\times}^b 
\Big( \textbf{S}_{1,a}^{m-1} \cdot \textbf{S}_{1,b}^{m} + \textbf{S}_{2,a}^{m-1} \cdot \textbf{S}_{2,b}^{m} \Big)  +J_{\times}^c 
\Big( \textbf{S}_{1,a}^m \cdot \textbf{S}_{1,b}^{m} + \textbf{S}_{2,a}^{m} \cdot \textbf{S}_{2,b}^{m} \Big) \bigg],
\end{split}
\end{equation}
\end{widetext}
where $m$ is the cell index, $a$ and $b$ denote the two legs of the ladder, the number $1$ or $2$ distinguishes the upper and the lower 
spin-site of a dimer and $\textbf{S}_{i,\alpha}^m$ with $\alpha = \{a,b\}$ and $i = \{1,2\}$, are the spin-$1/2$ operators.

At $W = 0$, the system consists of isolated dimers and the unperturbed ground state corresponds to a product of singlets on the 
leg dimers. The first excited state is the one-triplet state $\left| t \right\rangle_{\alpha}^m$, a state with a single triplet 
on a dimer $(m,\alpha)$ and singlets on all the other dimers. As Li$_2$Cu$_2$O(SO$_4$)$_2$ contains two dimers per unit cell, 
a $2\times2$ effective Hamiltonian, $W$ has to be computed for each value of $k$ in Fourier space. This leads to two separate bands of triplets. 
The dispersion relation is obtained by diagonalizing the effective Hamiltonian up to a given order starting from the Bloch states 
$\left| T \right\rangle_{\alpha} = \frac{2}{\sqrt{N_c}} \sum_m e^{i k m} \left| t \right\rangle_{\alpha}^m$, where $\alpha= \{ a, b\}$ and $N_c$ 
is the number of unit cells. 
\begin{figure}[hb!]
  \includegraphics[width=0.48 \textwidth]{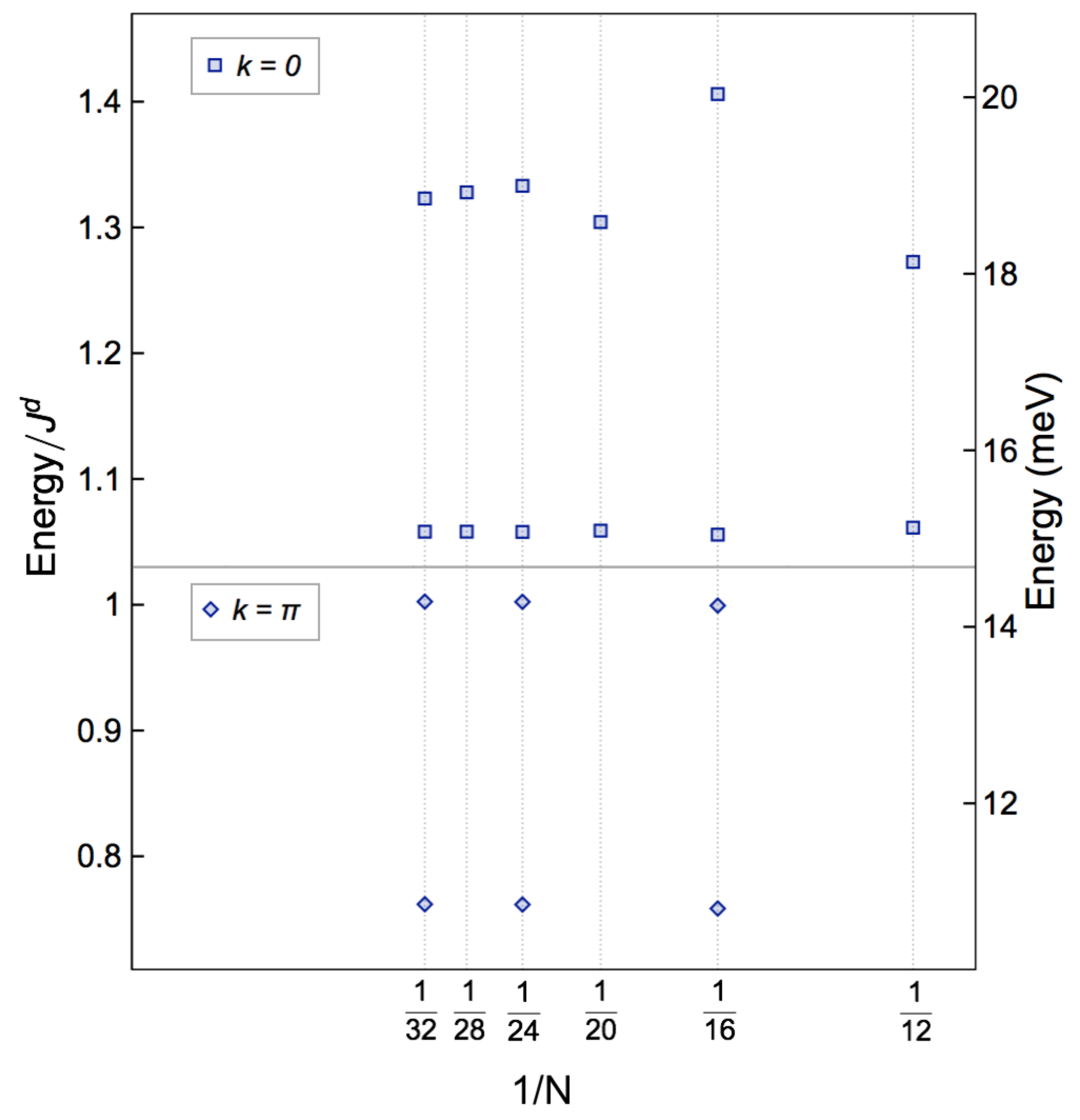}
  \caption{\label{fig:fig8} Exact diagonalization results obtained on finite lattices up to $N = 32$ sites.
 The triplet ($S = 1$) energies for $k = 0$ and $k = \pi$ are plotted as a function of $1/N$.}
\end{figure}

\subsection{First order}

At first order, the effective Hamiltonian ${\textbf{W}}_{\text{o}1}$
is given simply by the matrix elements of Eq.~(\ref{eq:PT_per_tri}) between
the states $\left| T \right\rangle_{\alpha}$.
The diagonal elements ${\textbf{W}}_{\text{o}1}(1,1) = {\textbf{W}}_{\text{o}1}(2,2)$
and the off-diagonal ${\textbf{W}}_{\text{o}1}(2,1) = {\textbf{W}}_{\text{o}1}^{*}(1,2)$ 
can be found by elementary means and read:
\begin{align}
{\textbf{W}}_{\text{o}1}(1, 1) &= \left( J_2 - \frac{J_a}{2} \right) \cos k, \\
\textbf{W}_{\text{o}1}(1, 2) &= \frac{1}{4} \Big( (\cos k - i \sin k ) ( 2 J_{\times}^b  - J_{\perp}^a) + 2J_{\times}^c - J_{\perp}^b \Big).
\end{align}

\subsection{Second order}

Computation of the second order is already more cumbersome, but can still be performed by elementary means (following for example the strategy 
of sections IV and V of Ref.~\onlinecite{Kogut79}). The second-order corrections ${\textbf{W}}_{\text{o}2}(1,1) = {\textbf{W}}_{\text{o}2}(2,2)$
and ${\textbf{W}}_{\text{o}2}(2,1) = {\textbf{W}}_{\text{o}2}^{*}(1,2)$ read
\begin{equation}
\begin{split}
{\textbf{W}}_{\text{o}2}&(1, 1) = \frac{1}{32} \Big( 12(J_{\times}^c - J_{\perp}^b)J_{\times}^c - (J_{\perp}^a)^2 - (J_{\perp}^b)^2 \\ 
& +  12(J_{\times}^b - J_{\perp}^a)J_{\times}^b - 2(J^a)^2 + 24(J_2 - J^a)J_2 \\ & - 2(2J_2 - J^a)^2 \cos (2k) + 
 2((2J_{\times}^c - J_{\perp}^b)J_{\perp}^a  \\& - 4(J^a)^2   - 2(2J_{\times}^c - J_{\perp}^b)J_{\times}^b) \cos k \Big),
\end{split}
\end{equation}

\begin{equation}
\begin{split}
{\textbf{W}}_{\text{o}2}&(1, 2) = \frac{1}{16} \Big( 2( - (J_{\perp}^b)^2 + i \sin k (J_{\perp}^a)^2 )  - (2J_2 \\ 
& - J^a)(2J_{\times}^b - J_{\perp}^a) - ( \cos (2k) - i \sin (2k) ) (2J_2 \\& - J^a) (2J_{\times}^b - J_{\perp}^a) - 
 2((2J_2 - J^a)(2J_{\times}^c - J_{\perp}^b) \\& + (J_{\perp}^a)^2) \cos k \Big).
\end{split}
\end{equation}

\subsection{Fifth order}

To proceed beyond second order, we have adapted a C-code that was used previously, e.g., in Refs.~\onlinecite{CHP98,perturb99} 
to the present situation. The main challenges here include the large number of variables (6 coupling ratios) as well as 
the rather long-range nature of some couplings (see Fig.~\ref{fig:fig1}(e)). This limits us to fifth order, 
the resulting expressions are too cumbersome to be presented here (the full expression of the perturbation matrix $W$ 
can be found in the Mathematica notebook provided in the Supplemental Material~\cite{SuppMat} that also contains a seventh-order
series for the ground-state energy). Nevertheless, we note 
that there is a freedom of the choice of basis and implementation of the perturbation theory that starts to impact the effective 
Hamiltonian $W$ at order three. For efficiency reasons, we decided to implement the perturbative corrections to the basis 
vectors starting from $\left| T \right\rangle_{\alpha}$ such that orthonormality is lost at higher order. 
As a consequence, the matrix representation of $W$ no longer needs to be Hermitian and, indeed, we find complex diagonal 
matrix elements starting at order 3. Nevertheless, we have checked that the eigenvalues of this $2 \times 2$ matrix are 
always real, as they should be since they are independent of the choice of basis.

We have also double-checked that we recover known limiting cases from our 6-variable expressions: 
for $J_\perp^a = J_\perp^b = J^a = J_2 = 0$ and $J_\times^b = J_\times^c$ we recover the conventional two-leg ladder, 
and our results for the eigenvalues are indeed consistent with Ref.~\onlinecite{Reigrotzki1994}; 
for $J_\perp^a = J_\perp^b = J_\times^b = J_\times^c = 0$ our model specializes to two decoupled frustrated and 
dimerized chains, a system that has been studied in Ref.~\onlinecite{Knetter2000} and  Ref.~\onlinecite{Barnes1999} for $J_2 = 0$.

For illustration purposes, we report a numerical expression of the dispersion relation up to full fifth order, 
obtained by substituting the DFT values of the coupling constants given in section \ref{sec:theo}
\begin{equation}
\label{equ:fifthorder}
\begin{split}
& \omega_{\text{o5} \pm} ( k ) = 1.09 + 1.65\cdot 10^{-1} \cos(k) - 3.54\cdot 10^{-2} \cos(2 k) \\ & 
\quad - 1.60\cdot 10^{-3} \cos(3 k) - 2.68\cdot 10^{-3} \cos(4 k) \\& 
\quad + 1.54\cdot 10^{-4} \cos(5k) \pm  \Big( 4.07\cdot 10^{-2} - 8.75\cdot 10^{-4} \cos(k) \\&
\quad - 2.35\cdot 10^{-2} \cos(2 k) + 4.38\cdot 10^{-3} \cos(3 k) \\ &
\quad + 7.68\cdot 10^{-4} \cos(4 k) + 1.94\cdot 10^{-4} \cos(5 k) \\ & 
\quad - 1.20\cdot 10^{-4} \cos(6 k) - 2.27 \cdot 10^{-6} \cos(7 k) \\& 
\quad + 1.04 \cdot 10^{-6} \cos(8 k) + 7.15 \cdot 10^{-7} \cos(9 k) \Big)^{1/2} .
\end{split}
\end{equation}

For reasons of compactness of presentation, we have rounded the numerical coefficients to three significant digits.
Expression (\ref{equ:fifthorder}) has been used in Fig.~\ref{fig:fig7} for order five in panel (a) and systematically in panel (b).

\section{Finite-size effects}
\label{appB}

Here we take a closer look at the finite-size effects present in the numerical results
for the one-triplet dispersion shown in section~\ref{sec:theo}. Indeed,
Fig.~\ref{fig:fig7}(b) exhibits a remarkable convergence of the numerical results towards
the fifth-order perturbative expansion with increasing $N$.
Figure~\ref{fig:fig8} shows the ED energies of the triplet excitations for two different values of $k$, 
$k=0$ and $k=\pi$, as a function of the inverse of the size, $1/N$, including the smaller system sizes
$N=12$ and $16$ not shown previously. One observes that
the values of the energies for $k = \pi$ and for the lower band of the triplet at $k = 0$  converge
rapidly. Larger finite-size effects are only observed at the top of the upper band at $k = \pi$.
This corresponds to the region where the series also show a slow convergence (see Fig.~\ref{fig:fig7}(a)) 
and we speculate that this is again due to the proximity with the continuum. Still, for systems with $N > 20$, the
data can also be considered to converge to the thermodynamic limit. Even in this least favorable case,
finite-size corrections to the $N=32$ data are presumably negligible for both bands and all values of $k$.

\bibliography{LiCuOSO_final}

\begin{thebibliography}{53}%
\makeatletter
\providecommand \@ifxundefined [1]{%
 \@ifx{#1\undefined}
}%
\providecommand \@ifnum [1]{%
 \ifnum #1\expandafter \@firstoftwo
 \else \expandafter \@secondoftwo
 \fi
}%
\providecommand \@ifx [1]{%
 \ifx #1\expandafter \@firstoftwo
 \else \expandafter \@secondoftwo
 \fi
}%
\providecommand \natexlab [1]{#1}%
\providecommand \enquote  [1]{``#1''}%
\providecommand \bibnamefont  [1]{#1}%
\providecommand \bibfnamefont [1]{#1}%
\providecommand \citenamefont [1]{#1}%
\providecommand \href@noop [0]{\@secondoftwo}%
\providecommand \href [0]{\begingroup \@sanitize@url \@href}%
\providecommand \@href[1]{\@@startlink{#1}\@@href}%
\providecommand \@@href[1]{\endgroup#1\@@endlink}%
\providecommand \@sanitize@url [0]{\catcode `\\12\catcode `\$12\catcode
  `\&12\catcode `\#12\catcode `\^12\catcode `\_12\catcode `\%12\relax}%
\providecommand \@@startlink[1]{}%
\providecommand \@@endlink[0]{}%
\providecommand \url  [0]{\begingroup\@sanitize@url \@url }%
\providecommand \@url [1]{\endgroup\@href {#1}{\urlprefix }}%
\providecommand \urlprefix  [0]{URL }%
\providecommand \Eprint [0]{\href }%
\providecommand \doibase [0]{http://dx.doi.org/}%
\providecommand \selectlanguage [0]{\@gobble}%
\providecommand \bibinfo  [0]{\@secondoftwo}%
\providecommand \bibfield  [0]{\@secondoftwo}%
\providecommand \translation [1]{[#1]}%
\providecommand \BibitemOpen [0]{}%
\providecommand \bibitemStop [0]{}%
\providecommand \bibitemNoStop [0]{.\EOS\space}%
\providecommand \EOS [0]{\spacefactor3000\relax}%
\providecommand \BibitemShut  [1]{\csname bibitem#1\endcsname}%
\let\auto@bib@innerbib\@empty
\bibitem [{\citenamefont {Lacroix}\ \emph {et~al.}(2011)\citenamefont
  {Lacroix}, \citenamefont {Mendels},\ and\ \citenamefont
  {Mila}}]{lacroix2011introduction}%
  \BibitemOpen
  \bibfield  {author} {\bibinfo {author} {\bibfnamefont {C.}~\bibnamefont
  {Lacroix}}, \bibinfo {author} {\bibfnamefont {P.}~\bibnamefont {Mendels}}, \
  and\ \bibinfo {author} {\bibfnamefont {F.}~\bibnamefont {Mila}},\ }\href@noop
  {} {\emph {\bibinfo {title} {Introduction to Frustrated Magnetism: Materials,
  Experiments, Theory}}},\ \bibinfo {series} {Springer Series in Solid-State
  Sciences}, Vol.\ \bibinfo {volume} {164}\ (\bibinfo  {publisher} {Springer
  Berlin Heidelberg},\ \bibinfo {year} {2011})\BibitemShut {NoStop}%
\bibitem [{\citenamefont {Mila}(1998)}]{Mila1998}%
  \BibitemOpen
  \bibfield  {author} {\bibinfo {author} {\bibfnamefont {F.}~\bibnamefont
  {Mila}},\ }\href@noop {} {\bibfield  {journal} {\bibinfo  {journal} {Eur.
  Phys. J. B}\ }\textbf {\bibinfo {volume} {6}},\ \bibinfo {pages} {201--205}
  (\bibinfo {year} {1998})}\BibitemShut {NoStop}%
\bibitem [{\citenamefont {Honecker}\ \emph {et~al.}(2000)\citenamefont
  {Honecker}, \citenamefont {Mila},\ and\ \citenamefont
  {Troyer}}]{Honecker2000}%
  \BibitemOpen
  \bibfield  {author} {\bibinfo {author} {\bibfnamefont {A.}~\bibnamefont
  {Honecker}}, \bibinfo {author} {\bibfnamefont {F.}~\bibnamefont {Mila}}, \
  and\ \bibinfo {author} {\bibfnamefont {M.}~\bibnamefont {Troyer}},\
  }\href@noop {} {\bibfield  {journal} {\bibinfo  {journal} {Eur. Phys. J. B}\
  }\textbf {\bibinfo {volume} {15}},\ \bibinfo {pages} {227--233} (\bibinfo
  {year} {2000})}\BibitemShut {NoStop}%
\bibitem [{\citenamefont {Vekua}\ and\ \citenamefont
  {Honecker}(2006)}]{Vekua2006}%
  \BibitemOpen
  \bibfield  {author} {\bibinfo {author} {\bibfnamefont {T.}~\bibnamefont
  {Vekua}}\ and\ \bibinfo {author} {\bibfnamefont {A.}~\bibnamefont
  {Honecker}},\ }\href@noop {} {\bibfield  {journal} {\bibinfo  {journal}
  {Phys. Rev. B}\ }\textbf {\bibinfo {volume} {73}},\ \bibinfo {pages} {214427}
  (\bibinfo {year} {2006})}\BibitemShut {NoStop}%
\bibitem [{\citenamefont {Liu}\ \emph {et~al.}(2008)\citenamefont {Liu},
  \citenamefont {Wang},\ and\ \citenamefont {Tian}}]{Liu2008}%
  \BibitemOpen
  \bibfield  {author} {\bibinfo {author} {\bibfnamefont {G.-H.}\ \bibnamefont
  {Liu}}, \bibinfo {author} {\bibfnamefont {H.-L.}\ \bibnamefont {Wang}}, \
  and\ \bibinfo {author} {\bibfnamefont {G.-S.}\ \bibnamefont {Tian}},\
  }\href@noop {} {\bibfield  {journal} {\bibinfo  {journal} {Phys. Rev. B}\
  }\textbf {\bibinfo {volume} {77}},\ \bibinfo {pages} {214418} (\bibinfo
  {year} {2008})}\BibitemShut {NoStop}%
\bibitem [{\citenamefont {Wen}\ \emph {et~al.}(2011)\citenamefont {Wen},
  \citenamefont {Liu},\ and\ \citenamefont {Tian}}]{RGHS2011}%
  \BibitemOpen
  \bibfield  {author} {\bibinfo {author} {\bibfnamefont {R.}~\bibnamefont
  {Wen}}, \bibinfo {author} {\bibfnamefont {G.-H.}\ \bibnamefont {Liu}}, \ and\
  \bibinfo {author} {\bibfnamefont {G.-S.}\ \bibnamefont {Tian}},\ }\href@noop
  {} {\bibfield  {journal} {\bibinfo  {journal} {Commun. Theor. Phys.}\
  }\textbf {\bibinfo {volume} {55}},\ \bibinfo {pages} {1102--1108} (\bibinfo
  {year} {2011})}\BibitemShut {NoStop}%
\bibitem [{\citenamefont {Li}\ and\ \citenamefont {Lin}(2012)}]{LiLin2012}%
  \BibitemOpen
  \bibfield  {author} {\bibinfo {author} {\bibfnamefont {Y.-C.}\ \bibnamefont
  {Li}}\ and\ \bibinfo {author} {\bibfnamefont {H.-Q.}\ \bibnamefont {Lin}},\
  }\href@noop {} {\bibfield  {journal} {\bibinfo  {journal} {New J. Phys.}\
  }\textbf {\bibinfo {volume} {14}},\ \bibinfo {pages} {063019} (\bibinfo
  {year} {2012})}\BibitemShut {NoStop}%
\bibitem [{\citenamefont {Tsirlin}\ \emph {et~al.}(2010)\citenamefont
  {Tsirlin}, \citenamefont {Rousochatzakis}, \citenamefont {Kasinathan},
  \citenamefont {Janson}, \citenamefont {Nath}, \citenamefont {Weickert},
  \citenamefont {Geibel}, \citenamefont {L\"auchli},\ and\ \citenamefont
  {Rosner}}]{Tsirlin2010}%
  \BibitemOpen
  \bibfield  {author} {\bibinfo {author} {\bibfnamefont {A.~A.}\ \bibnamefont
  {Tsirlin}}, \bibinfo {author} {\bibfnamefont {I.}~\bibnamefont
  {Rousochatzakis}}, \bibinfo {author} {\bibfnamefont {D.}~\bibnamefont
  {Kasinathan}}, \bibinfo {author} {\bibfnamefont {O.}~\bibnamefont {Janson}},
  \bibinfo {author} {\bibfnamefont {R.}~\bibnamefont {Nath}}, \bibinfo {author}
  {\bibfnamefont {F.}~\bibnamefont {Weickert}}, \bibinfo {author}
  {\bibfnamefont {C.}~\bibnamefont {Geibel}}, \bibinfo {author} {\bibfnamefont
  {A.~M.}\ \bibnamefont {L\"auchli}}, \ and\ \bibinfo {author} {\bibfnamefont
  {H.}~\bibnamefont {Rosner}},\ }\href@noop {} {\bibfield  {journal} {\bibinfo
  {journal} {Phys. Rev. B}\ }\textbf {\bibinfo {volume} {82}},\ \bibinfo
  {pages} {144426} (\bibinfo {year} {2010})}\BibitemShut {NoStop}%
\bibitem [{\citenamefont {Koteswararao}\ \emph {et~al.}(2007)\citenamefont
  {Koteswararao}, \citenamefont {Salunke}, \citenamefont {Mahajan},
  \citenamefont {Dasgupta},\ and\ \citenamefont {Bobroff}}]{Koteswararao2007}%
  \BibitemOpen
  \bibfield  {author} {\bibinfo {author} {\bibfnamefont {B.}~\bibnamefont
  {Koteswararao}}, \bibinfo {author} {\bibfnamefont {S.}~\bibnamefont
  {Salunke}}, \bibinfo {author} {\bibfnamefont {A.~V.}\ \bibnamefont
  {Mahajan}}, \bibinfo {author} {\bibfnamefont {I.}~\bibnamefont {Dasgupta}}, \
  and\ \bibinfo {author} {\bibfnamefont {J.}~\bibnamefont {Bobroff}},\
  }\href@noop {} {\bibfield  {journal} {\bibinfo  {journal} {Phys. Rev. B}\
  }\textbf {\bibinfo {volume} {76}},\ \bibinfo {pages} {052402} (\bibinfo
  {year} {2007})}\BibitemShut {NoStop}%
\bibitem [{\citenamefont {Mentr\'e}\ \emph {et~al.}(2009)\citenamefont
  {Mentr\'e}, \citenamefont {Janod}, \citenamefont {Rabu}, \citenamefont
  {Hennion}, \citenamefont {Leclercq-Hugeux}, \citenamefont {Kang},
  \citenamefont {Lee}, \citenamefont {Whangbo},\ and\ \citenamefont
  {Petit}}]{Mentre2009}%
  \BibitemOpen
  \bibfield  {author} {\bibinfo {author} {\bibfnamefont {O.}~\bibnamefont
  {Mentr\'e}}, \bibinfo {author} {\bibfnamefont {E.}~\bibnamefont {Janod}},
  \bibinfo {author} {\bibfnamefont {P.}~\bibnamefont {Rabu}}, \bibinfo {author}
  {\bibfnamefont {M.}~\bibnamefont {Hennion}}, \bibinfo {author} {\bibfnamefont
  {F.}~\bibnamefont {Leclercq-Hugeux}}, \bibinfo {author} {\bibfnamefont
  {J.}~\bibnamefont {Kang}}, \bibinfo {author} {\bibfnamefont {C.}~\bibnamefont
  {Lee}}, \bibinfo {author} {\bibfnamefont {M.-H.}\ \bibnamefont {Whangbo}}, \
  and\ \bibinfo {author} {\bibfnamefont {S.}~\bibnamefont {Petit}},\
  }\href@noop {} {\bibfield  {journal} {\bibinfo  {journal} {Phys. Rev. B}\
  }\textbf {\bibinfo {volume} {80}},\ \bibinfo {pages} {180413} (\bibinfo
  {year} {2009})}\BibitemShut {NoStop}%
\bibitem [{\citenamefont {Plumb}\ \emph {et~al.}(2013)\citenamefont {Plumb},
  \citenamefont {Yamani}, \citenamefont {Matsuda}, \citenamefont {Shu},
  \citenamefont {Koteswararao}, \citenamefont {Chou},\ and\ \citenamefont
  {Kim}}]{Plumb2013}%
  \BibitemOpen
  \bibfield  {author} {\bibinfo {author} {\bibfnamefont {K.~W.}\ \bibnamefont
  {Plumb}}, \bibinfo {author} {\bibfnamefont {Z.}~\bibnamefont {Yamani}},
  \bibinfo {author} {\bibfnamefont {M.}~\bibnamefont {Matsuda}}, \bibinfo
  {author} {\bibfnamefont {G.~J.}\ \bibnamefont {Shu}}, \bibinfo {author}
  {\bibfnamefont {B.}~\bibnamefont {Koteswararao}}, \bibinfo {author}
  {\bibfnamefont {F.~C.}\ \bibnamefont {Chou}}, \ and\ \bibinfo {author}
  {\bibfnamefont {Y.-J.}\ \bibnamefont {Kim}},\ }\href@noop {} {\bibfield
  {journal} {\bibinfo  {journal} {Phys. Rev. B}\ }\textbf {\bibinfo {volume}
  {88}},\ \bibinfo {pages} {024402} (\bibinfo {year} {2013})}\BibitemShut
  {NoStop}%
\bibitem [{\citenamefont {Plumb}\ \emph {et~al.}(2016)\citenamefont {Plumb},
  \citenamefont {Hwang}, \citenamefont {Qiu}, \citenamefont {Harriger},
  \citenamefont {Granroth}, \citenamefont {Kolesnikov}, \citenamefont {Shu},
  \citenamefont {Chou}, \citenamefont {R\"uegg}, \citenamefont {Kim},\ and\
  \citenamefont {Kim}}]{Plumb2016}%
  \BibitemOpen
  \bibfield  {author} {\bibinfo {author} {\bibfnamefont {K.~W.}\ \bibnamefont
  {Plumb}}, \bibinfo {author} {\bibfnamefont {K.}~\bibnamefont {Hwang}},
  \bibinfo {author} {\bibfnamefont {Y.}~\bibnamefont {Qiu}}, \bibinfo {author}
  {\bibfnamefont {L.~W.}\ \bibnamefont {Harriger}}, \bibinfo {author}
  {\bibfnamefont {G.~E.}\ \bibnamefont {Granroth}}, \bibinfo {author}
  {\bibfnamefont {A.~I.}\ \bibnamefont {Kolesnikov}}, \bibinfo {author}
  {\bibfnamefont {G.~J.}\ \bibnamefont {Shu}}, \bibinfo {author} {\bibfnamefont
  {F.~C.}\ \bibnamefont {Chou}}, \bibinfo {author} {\bibfnamefont
  {C.}~\bibnamefont {R\"uegg}}, \bibinfo {author} {\bibfnamefont {Y.~B.}\
  \bibnamefont {Kim}}, \ and\ \bibinfo {author} {\bibfnamefont {Y.-J.}\
  \bibnamefont {Kim}},\ }\href@noop {} {\bibfield  {journal} {\bibinfo
  {journal} {Nat. Phys.}\ }\textbf {\bibinfo {volume} {12}},\ \bibinfo {pages}
  {224--229} (\bibinfo {year} {2016})}\BibitemShut {NoStop}%
\bibitem [{\citenamefont {Kohama}\ \emph {et~al.}(2012)\citenamefont {Kohama},
  \citenamefont {Wang}, \citenamefont {Uchida}, \citenamefont {Prsa},
  \citenamefont {Zvyagin}, \citenamefont {Skourski}, \citenamefont {McDonald},
  \citenamefont {Balicas}, \citenamefont {Ronnow}, \citenamefont {R\"uegg},\
  and\ \citenamefont {Jaime}}]{Kohama2012}%
  \BibitemOpen
  \bibfield  {author} {\bibinfo {author} {\bibfnamefont {Y.}~\bibnamefont
  {Kohama}}, \bibinfo {author} {\bibfnamefont {S.}~\bibnamefont {Wang}},
  \bibinfo {author} {\bibfnamefont {A.}~\bibnamefont {Uchida}}, \bibinfo
  {author} {\bibfnamefont {K.}~\bibnamefont {Prsa}}, \bibinfo {author}
  {\bibfnamefont {S.}~\bibnamefont {Zvyagin}}, \bibinfo {author} {\bibfnamefont
  {Y.}~\bibnamefont {Skourski}}, \bibinfo {author} {\bibfnamefont {R.~D.}\
  \bibnamefont {McDonald}}, \bibinfo {author} {\bibfnamefont {L.}~\bibnamefont
  {Balicas}}, \bibinfo {author} {\bibfnamefont {H.~M.}\ \bibnamefont {Ronnow}},
  \bibinfo {author} {\bibfnamefont {C.}~\bibnamefont {R\"uegg}}, \ and\
  \bibinfo {author} {\bibfnamefont {M.}~\bibnamefont {Jaime}},\ }\href@noop {}
  {\bibfield  {journal} {\bibinfo  {journal} {Phys. Rev. Lett.}\ }\textbf
  {\bibinfo {volume} {109}},\ \bibinfo {pages} {167204} (\bibinfo {year}
  {2012})}\BibitemShut {NoStop}%
\bibitem [{\citenamefont {Casola}\ \emph {et~al.}(2013)\citenamefont {Casola},
  \citenamefont {Shiroka}, \citenamefont {Feiguin}, \citenamefont {Wang},
  \citenamefont {Grbi\ifmmode~\acute{c}\else \'{c}\fi{}}, \citenamefont
  {Horvati\ifmmode~\acute{c}\else \'{c}\fi{}}, \citenamefont {Kr\"amer},
  \citenamefont {Mukhopadhyay}, \citenamefont {Conder}, \citenamefont
  {Berthier}, \citenamefont {Ott}, \citenamefont {R\o{}nnow}, \citenamefont
  {R\"uegg},\ and\ \citenamefont {Mesot}}]{Casola2013}%
  \BibitemOpen
  \bibfield  {author} {\bibinfo {author} {\bibfnamefont {F.}~\bibnamefont
  {Casola}}, \bibinfo {author} {\bibfnamefont {T.}~\bibnamefont {Shiroka}},
  \bibinfo {author} {\bibfnamefont {A.}~\bibnamefont {Feiguin}}, \bibinfo
  {author} {\bibfnamefont {S.}~\bibnamefont {Wang}}, \bibinfo {author}
  {\bibfnamefont {M.~S.}\ \bibnamefont {Grbi\ifmmode~\acute{c}\else
  \'{c}\fi{}}}, \bibinfo {author} {\bibfnamefont {M.}~\bibnamefont
  {Horvati\ifmmode~\acute{c}\else \'{c}\fi{}}}, \bibinfo {author}
  {\bibfnamefont {S.}~\bibnamefont {Kr\"amer}}, \bibinfo {author}
  {\bibfnamefont {S.}~\bibnamefont {Mukhopadhyay}}, \bibinfo {author}
  {\bibfnamefont {K.}~\bibnamefont {Conder}}, \bibinfo {author} {\bibfnamefont
  {C.}~\bibnamefont {Berthier}}, \bibinfo {author} {\bibfnamefont {H.-R.}\
  \bibnamefont {Ott}}, \bibinfo {author} {\bibfnamefont {H.~M.}\ \bibnamefont
  {R\o{}nnow}}, \bibinfo {author} {\bibfnamefont {Ch.}\ \bibnamefont
  {R\"uegg}}, \ and\ \bibinfo {author} {\bibfnamefont {J.}~\bibnamefont
  {Mesot}},\ }\href@noop {} {\bibfield  {journal} {\bibinfo  {journal} {Phys.
  Rev. Lett.}\ }\textbf {\bibinfo {volume} {110}},\ \bibinfo {pages} {187201}
  (\bibinfo {year} {2013})}\BibitemShut {NoStop}%
\bibitem [{\citenamefont {Kohama}\ \emph {et~al.}(2014)\citenamefont {Kohama},
  \citenamefont {Mochidzuki}, \citenamefont {Terashima}, \citenamefont
  {Miyata}, \citenamefont {DeMuer}, \citenamefont {Klein}, \citenamefont
  {Marcenat}, \citenamefont {Dun}, \citenamefont {Zhou}, \citenamefont {Li},
  \citenamefont {Balicas}, \citenamefont {Abe}, \citenamefont {Matsuda},
  \citenamefont {Takeyama}, \citenamefont {Matsuo},\ and\ \citenamefont
  {Kindo}}]{Kohama2014}%
  \BibitemOpen
  \bibfield  {author} {\bibinfo {author} {\bibfnamefont {Y.}~\bibnamefont
  {Kohama}}, \bibinfo {author} {\bibfnamefont {K.}~\bibnamefont {Mochidzuki}},
  \bibinfo {author} {\bibfnamefont {T.}~\bibnamefont {Terashima}}, \bibinfo
  {author} {\bibfnamefont {A.}~\bibnamefont {Miyata}}, \bibinfo {author}
  {\bibfnamefont {A.}~\bibnamefont {DeMuer}}, \bibinfo {author} {\bibfnamefont
  {T.}~\bibnamefont {Klein}}, \bibinfo {author} {\bibfnamefont
  {C.}~\bibnamefont {Marcenat}}, \bibinfo {author} {\bibfnamefont {Z.~L.}\
  \bibnamefont {Dun}}, \bibinfo {author} {\bibfnamefont {H.}~\bibnamefont
  {Zhou}}, \bibinfo {author} {\bibfnamefont {G.}~\bibnamefont {Li}}, \bibinfo
  {author} {\bibfnamefont {L.}~\bibnamefont {Balicas}}, \bibinfo {author}
  {\bibfnamefont {N.}~\bibnamefont {Abe}}, \bibinfo {author} {\bibfnamefont
  {Y.~H.}\ \bibnamefont {Matsuda}}, \bibinfo {author} {\bibfnamefont
  {S.}~\bibnamefont {Takeyama}}, \bibinfo {author} {\bibfnamefont
  {A.}~\bibnamefont {Matsuo}}, \ and\ \bibinfo {author} {\bibfnamefont
  {K.}~\bibnamefont {Kindo}},\ }\href@noop {} {\bibfield  {journal} {\bibinfo
  {journal} {Phys. Rev. B}\ }\textbf {\bibinfo {volume} {90}},\ \bibinfo
  {pages} {060408} (\bibinfo {year} {2014})}\BibitemShut {NoStop}%
\bibitem [{\citenamefont {Sun}\ \emph {et~al.}(2015)\citenamefont {Sun},
  \citenamefont {Rousse}, \citenamefont {Abakumov}, \citenamefont
  {Sauban\`ere}, \citenamefont {Doublet}, \citenamefont
  {Rodr\'{\i}guez-Carvajal}, \citenamefont {Van~Tendeloo},\ and\ \citenamefont
  {Tarascon}}]{Sun2015}%
  \BibitemOpen
  \bibfield  {author} {\bibinfo {author} {\bibfnamefont {M.}~\bibnamefont
  {Sun}}, \bibinfo {author} {\bibfnamefont {G.}~\bibnamefont {Rousse}},
  \bibinfo {author} {\bibfnamefont {A.~M.}\ \bibnamefont {Abakumov}}, \bibinfo
  {author} {\bibfnamefont {M.}~\bibnamefont {Sauban\`ere}}, \bibinfo {author}
  {\bibfnamefont {M.-L.}\ \bibnamefont {Doublet}}, \bibinfo {author}
  {\bibfnamefont {J.}~\bibnamefont {Rodr\'{\i}guez-Carvajal}}, \bibinfo
  {author} {\bibfnamefont {G.}~\bibnamefont {Van~Tendeloo}}, \ and\ \bibinfo
  {author} {\bibfnamefont {J.-M.}\ \bibnamefont {Tarascon}},\ }\href@noop {}
  {\bibfield  {journal} {\bibinfo  {journal} {Chem. Mater.}\ }\textbf {\bibinfo
  {volume} {27}},\ \bibinfo {pages} {3077--3087} (\bibinfo {year}
  {2015})}\BibitemShut {NoStop}%
\bibitem [{\citenamefont {Rousse}\ \emph {et~al.}(2017)\citenamefont {Rousse},
  \citenamefont {Rodr\'{\i}guez-Carvajal}, \citenamefont {Giacobbe},
  \citenamefont {Sun}, \citenamefont {Vaccarelli},\ and\ \citenamefont
  {Radtke}}]{Rousse2017}%
  \BibitemOpen
  \bibfield  {author} {\bibinfo {author} {\bibfnamefont {G.}~\bibnamefont
  {Rousse}}, \bibinfo {author} {\bibfnamefont {J.}~\bibnamefont
  {Rodr\'{\i}guez-Carvajal}}, \bibinfo {author} {\bibfnamefont
  {C.}~\bibnamefont {Giacobbe}}, \bibinfo {author} {\bibfnamefont
  {M.}~\bibnamefont {Sun}}, \bibinfo {author} {\bibfnamefont {O.}~\bibnamefont
  {Vaccarelli}}, \ and\ \bibinfo {author} {\bibfnamefont {G.}~\bibnamefont
  {Radtke}},\ }\href@noop {} {\bibfield  {journal} {\bibinfo  {journal} {Phys.
  Rev. B}\ }\textbf {\bibinfo {volume} {95}},\ \bibinfo {pages} {144103}
  (\bibinfo {year} {2017})}\BibitemShut {NoStop}%
\bibitem [{\citenamefont {Vaccarelli}\ \emph {et~al.}(2017)\citenamefont
  {Vaccarelli}, \citenamefont {Rousse}, \citenamefont {Sa\'ul},\ and\
  \citenamefont {Radtke}}]{Vaccarelli2017}%
  \BibitemOpen
  \bibfield  {author} {\bibinfo {author} {\bibfnamefont {O.}~\bibnamefont
  {Vaccarelli}}, \bibinfo {author} {\bibfnamefont {G.}~\bibnamefont {Rousse}},
  \bibinfo {author} {\bibfnamefont {A.}~\bibnamefont {Sa\'ul}}, \ and\ \bibinfo
  {author} {\bibfnamefont {G.}~\bibnamefont {Radtke}},\ }\href@noop {}
  {\bibfield  {journal} {\bibinfo  {journal} {Phys. Rev. B}\ }\textbf {\bibinfo
  {volume} {96}},\ \bibinfo {pages} {180406} (\bibinfo {year}
  {2017})}\BibitemShut {NoStop}%
\bibitem [{\citenamefont {Richard}\ \emph {et~al.}(1996)\citenamefont
  {Richard}, \citenamefont {Ferrand},\ and\ \citenamefont
  {Kearley}}]{LAMP1996}%
  \BibitemOpen
  \bibfield  {author} {\bibinfo {author} {\bibfnamefont {D.}~\bibnamefont
  {Richard}}, \bibinfo {author} {\bibfnamefont {M.}~\bibnamefont {Ferrand}}, \
  and\ \bibinfo {author} {\bibfnamefont {G.J.}\ \bibnamefont {Kearley}},\
  }\href@noop {} {\bibfield  {journal} {\bibinfo  {journal} {J. Neutron
  Research}\ }\textbf {\bibinfo {volume} {4}},\ \bibinfo {pages} {33--39}
  (\bibinfo {year} {1996})}\BibitemShut {NoStop}%
\bibitem [{\citenamefont {Lovesey}(1984)}]{Lovesey1984}%
  \BibitemOpen
  \bibfield  {author} {\bibinfo {author} {\bibfnamefont {S.~W.}\ \bibnamefont
  {Lovesey}},\ }\href@noop {} {\emph {\bibinfo {title} {Theory of Neutron
  Scattering from Condensed Matter}}}\ (\bibinfo  {publisher} {Clarendon Press,
  Oxford UK},\ \bibinfo {year} {1984})\BibitemShut {NoStop}%
\bibitem [{\citenamefont {Squires}(1978)}]{Squires1978}%
  \BibitemOpen
  \bibfield  {author} {\bibinfo {author} {\bibfnamefont {G.~L.}\ \bibnamefont
  {Squires}},\ }\href@noop {} {\emph {\bibinfo {title} {Introduction to the
  Theory of Thermal Neutron Scattering}}}\ (\bibinfo  {publisher} {Dover
  Publications},\ \bibinfo {year} {1978})\BibitemShut {NoStop}%
\bibitem [{\citenamefont {Clancy}\ \emph {et~al.}(2011)\citenamefont {Clancy},
  \citenamefont {Gaulin}, \citenamefont {Adams}, \citenamefont {Granroth},
  \citenamefont {Kolesnikov}, \citenamefont {Sherline},\ and\ \citenamefont
  {Chou}}]{Clancy2011}%
  \BibitemOpen
  \bibfield  {author} {\bibinfo {author} {\bibfnamefont {J.~P.}\ \bibnamefont
  {Clancy}}, \bibinfo {author} {\bibfnamefont {B.~D.}\ \bibnamefont {Gaulin}},
  \bibinfo {author} {\bibfnamefont {C.~P.}\ \bibnamefont {Adams}}, \bibinfo
  {author} {\bibfnamefont {G.~E.}\ \bibnamefont {Granroth}}, \bibinfo {author}
  {\bibfnamefont {A.~I.}\ \bibnamefont {Kolesnikov}}, \bibinfo {author}
  {\bibfnamefont {T.~E.}\ \bibnamefont {Sherline}}, \ and\ \bibinfo {author}
  {\bibfnamefont {F.~C.}\ \bibnamefont {Chou}},\ }\href@noop {} {\bibfield
  {journal} {\bibinfo  {journal} {Phys. Rev. Lett.}\ }\textbf {\bibinfo
  {volume} {106}},\ \bibinfo {pages} {117401} (\bibinfo {year}
  {2011})}\BibitemShut {NoStop}%
\bibitem [{\citenamefont {Troyer}\ \emph {et~al.}(1994)\citenamefont {Troyer},
  \citenamefont {Tsunetsugu},\ and\ \citenamefont {W\"urtz}}]{Troyer1994}%
  \BibitemOpen
  \bibfield  {author} {\bibinfo {author} {\bibfnamefont {M.}~\bibnamefont
  {Troyer}}, \bibinfo {author} {\bibfnamefont {H.}~\bibnamefont {Tsunetsugu}},
  \ and\ \bibinfo {author} {\bibfnamefont {D.}~\bibnamefont {W\"urtz}},\
  }\href@noop {} {\bibfield  {journal} {\bibinfo  {journal} {Phys. Rev. B}\
  }\textbf {\bibinfo {volume} {50}},\ \bibinfo {pages} {13515--13527} (\bibinfo
  {year} {1994})}\BibitemShut {NoStop}%
\bibitem [{\citenamefont {Farmer}(1974)}]{Farmer1974}%
  \BibitemOpen
  \bibfield  {author} {\bibinfo {author} {\bibfnamefont {V.~C.}\ \bibnamefont
  {Farmer}},\ }\href@noop {} {\emph {\bibinfo {title} {The Infrared Spectra of
  Minerals}}}\ (\bibinfo  {publisher} {Mineralogical Society of Great Britain
  and Ireland},\ \bibinfo {year} {1974})\BibitemShut {NoStop}%
\bibitem [{\citenamefont {{R\~o\~om}}\ \emph
  {et~al.}(2004{\natexlab{a}})\citenamefont {{R\~o\~om}}, \citenamefont
  {H\"uvonen}, \citenamefont {Nagel}, \citenamefont {Hwang}, \citenamefont
  {Timusk},\ and\ \citenamefont {Kageyama}}]{Room2004b}%
  \BibitemOpen
  \bibfield  {author} {\bibinfo {author} {\bibfnamefont {T.}~\bibnamefont
  {{R\~o\~om}}}, \bibinfo {author} {\bibfnamefont {D.}~\bibnamefont
  {H\"uvonen}}, \bibinfo {author} {\bibfnamefont {U.}~\bibnamefont {Nagel}},
  \bibinfo {author} {\bibfnamefont {J.}~\bibnamefont {Hwang}}, \bibinfo
  {author} {\bibfnamefont {T.}~\bibnamefont {Timusk}}, \ and\ \bibinfo {author}
  {\bibfnamefont {H.}~\bibnamefont {Kageyama}},\ }\href@noop {} {\bibfield
  {journal} {\bibinfo  {journal} {Phys. Rev. B}\ }\textbf {\bibinfo {volume}
  {70}},\ \bibinfo {pages} {144417} (\bibinfo {year}
  {2004}{\natexlab{a}})}\BibitemShut {NoStop}%
\bibitem [{\citenamefont {Furrer}\ and\ \citenamefont
  {G\"udel}(1979)}]{Furrer1979}%
  \BibitemOpen
  \bibfield  {author} {\bibinfo {author} {\bibfnamefont {A.}~\bibnamefont
  {Furrer}}\ and\ \bibinfo {author} {\bibfnamefont {H.~U.}\ \bibnamefont
  {G\"udel}},\ }\href@noop {} {\bibfield  {journal} {\bibinfo  {journal} {J.
  Magn. Mag. Mat.}\ }\textbf {\bibinfo {volume} {14}},\ \bibinfo {pages} {256}
  (\bibinfo {year} {1979})}\BibitemShut {NoStop}%
\bibitem [{\citenamefont {Hohenberg}\ and\ \citenamefont
  {Brinkmann}(1974)}]{Hohenberg1974}%
  \BibitemOpen
  \bibfield  {author} {\bibinfo {author} {\bibfnamefont {P.~C.}\ \bibnamefont
  {Hohenberg}}\ and\ \bibinfo {author} {\bibfnamefont {W.~F.}\ \bibnamefont
  {Brinkmann}},\ }\href@noop {} {\bibfield  {journal} {\bibinfo  {journal}
  {Phys. Rev. B}\ }\textbf {\bibinfo {volume} {10}},\ \bibinfo {pages} {128}
  (\bibinfo {year} {1974})}\BibitemShut {NoStop}%
\bibitem [{\citenamefont {Stone}\ \emph {et~al.}(2002)\citenamefont {Stone},
  \citenamefont {Chen}, \citenamefont {Rittner}, \citenamefont {Yardimci},
  \citenamefont {Reich}, \citenamefont {Broholm}, \citenamefont {Ferraris},\
  and\ \citenamefont {Lectka}}]{Stone2002}%
  \BibitemOpen
  \bibfield  {author} {\bibinfo {author} {\bibfnamefont {M.~B.}\ \bibnamefont
  {Stone}}, \bibinfo {author} {\bibfnamefont {Y.}~\bibnamefont {Chen}},
  \bibinfo {author} {\bibfnamefont {J.}~\bibnamefont {Rittner}}, \bibinfo
  {author} {\bibfnamefont {H.}~\bibnamefont {Yardimci}}, \bibinfo {author}
  {\bibfnamefont {D.~H.}\ \bibnamefont {Reich}}, \bibinfo {author}
  {\bibfnamefont {C.}~\bibnamefont {Broholm}}, \bibinfo {author} {\bibfnamefont
  {D.~V.}\ \bibnamefont {Ferraris}}, \ and\ \bibinfo {author} {\bibfnamefont
  {T.}~\bibnamefont {Lectka}},\ }\href@noop {} {\bibfield  {journal} {\bibinfo
  {journal} {Phys. Rev. B}\ }\textbf {\bibinfo {volume} {65}},\ \bibinfo
  {pages} {064423} (\bibinfo {year} {2002})}\BibitemShut {NoStop}%
\bibitem [{\citenamefont {Quintero-Castro}\ \emph {et~al.}(2010)\citenamefont
  {Quintero-Castro}, \citenamefont {Lake}, \citenamefont {Wheller},
  \citenamefont {Islam}, \citenamefont {Guidi}, \citenamefont {Rule},
  \citenamefont {Izaola}, \citenamefont {Russina}, \citenamefont {Kiefer},\
  and\ \citenamefont {Skourski}}]{Quintero-Castro2010}%
  \BibitemOpen
  \bibfield  {author} {\bibinfo {author} {\bibfnamefont {D.~L.}\ \bibnamefont
  {Quintero-Castro}}, \bibinfo {author} {\bibfnamefont {B.}~\bibnamefont
  {Lake}}, \bibinfo {author} {\bibfnamefont {E.~M.}\ \bibnamefont {Wheller}},
  \bibinfo {author} {\bibfnamefont {A.~T. M.~N.}\ \bibnamefont {Islam}},
  \bibinfo {author} {\bibfnamefont {T.}~\bibnamefont {Guidi}}, \bibinfo
  {author} {\bibfnamefont {K.~C.}\ \bibnamefont {Rule}}, \bibinfo {author}
  {\bibfnamefont {Z.}~\bibnamefont {Izaola}}, \bibinfo {author} {\bibfnamefont
  {M.}~\bibnamefont {Russina}}, \bibinfo {author} {\bibfnamefont
  {K.}~\bibnamefont {Kiefer}}, \ and\ \bibinfo {author} {\bibfnamefont
  {Y.}~\bibnamefont {Skourski}},\ }\href@noop {} {\bibfield  {journal}
  {\bibinfo  {journal} {Phys. Rev. B}\ }\textbf {\bibinfo {volume} {81}},\
  \bibinfo {pages} {014415} (\bibinfo {year} {2010})}\BibitemShut {NoStop}%
\bibitem [{\citenamefont {Can\'evet}\ \emph {et~al.}(2015)\citenamefont
  {Can\'evet}, \citenamefont {F{\aa}k}, \citenamefont {Kremer}, \citenamefont
  {Chun}, \citenamefont {Enderle}, \citenamefont {Gordon}, \citenamefont
  {Bettis}, \citenamefont {Whangbo}, \citenamefont {Taylor},\ and\
  \citenamefont {Adroja}}]{Canevet2015}%
  \BibitemOpen
  \bibfield  {author} {\bibinfo {author} {\bibfnamefont {E.}~\bibnamefont
  {Can\'evet}}, \bibinfo {author} {\bibfnamefont {B.}~\bibnamefont {F{\aa}k}},
  \bibinfo {author} {\bibfnamefont {R.~K.}\ \bibnamefont {Kremer}}, \bibinfo
  {author} {\bibfnamefont {J.~H.}\ \bibnamefont {Chun}}, \bibinfo {author}
  {\bibfnamefont {M.}~\bibnamefont {Enderle}}, \bibinfo {author} {\bibfnamefont
  {E.~E.}\ \bibnamefont {Gordon}}, \bibinfo {author} {\bibfnamefont {J.~L.}\
  \bibnamefont {Bettis}}, \bibinfo {author} {\bibfnamefont {M.-H.}\
  \bibnamefont {Whangbo}}, \bibinfo {author} {\bibfnamefont {J.~W.}\
  \bibnamefont {Taylor}}, \ and\ \bibinfo {author} {\bibfnamefont {D.~T.}\
  \bibnamefont {Adroja}},\ }\href@noop {} {\bibfield  {journal} {\bibinfo
  {journal} {Phys. Rev. B}\ }\textbf {\bibinfo {volume} {91}},\ \bibinfo
  {pages} {060402(R)} (\bibinfo {year} {2015})}\BibitemShut {NoStop}%
\bibitem [{\citenamefont {Lanczos}(1950)}]{Lanczos1950}%
  \BibitemOpen
  \bibfield  {author} {\bibinfo {author} {\bibfnamefont {C.}~\bibnamefont
  {Lanczos}},\ }\href@noop {} {\bibfield  {journal} {\bibinfo  {journal} {J.
  Res. Nat. Bur. Stand.}\ }\textbf {\bibinfo {volume} {45}},\ \bibinfo {pages}
  {255} (\bibinfo {year} {1950})}\BibitemShut {NoStop}%
\bibitem [{\citenamefont {Dagotto}(1994)}]{Dagotto1994}%
  \BibitemOpen
  \bibfield  {author} {\bibinfo {author} {\bibfnamefont {E.}~\bibnamefont
  {Dagotto}},\ }\href@noop {} {\bibfield  {journal} {\bibinfo  {journal} {Rev.
  Mod. Phys.}\ }\textbf {\bibinfo {volume} {66}},\ \bibinfo {pages} {763}
  (\bibinfo {year} {1994})}\BibitemShut {NoStop}%
\bibitem [{\citenamefont {Trebst}\ \emph {et~al.}(2000)\citenamefont {Trebst},
  \citenamefont {Monien}, \citenamefont {Hamer}, \citenamefont {Weihong},\ and\
  \citenamefont {Singh}}]{Trebst2000}%
  \BibitemOpen
  \bibfield  {author} {\bibinfo {author} {\bibfnamefont {S.}~\bibnamefont
  {Trebst}}, \bibinfo {author} {\bibfnamefont {H.}~\bibnamefont {Monien}},
  \bibinfo {author} {\bibfnamefont {C.~J.}\ \bibnamefont {Hamer}}, \bibinfo
  {author} {\bibfnamefont {Z.}~\bibnamefont {Weihong}}, \ and\ \bibinfo
  {author} {\bibfnamefont {R.~R.~P.}\ \bibnamefont {Singh}},\ }\href@noop {}
  {\bibfield  {journal} {\bibinfo  {journal} {Phys. Rev. Lett.}\ }\textbf
  {\bibinfo {volume} {85}},\ \bibinfo {pages} {4373--4376} (\bibinfo {year}
  {2000})}\BibitemShut {NoStop}%
\bibitem [{\citenamefont {Windt}\ \emph {et~al.}(2001)\citenamefont {Windt},
  \citenamefont {Gr\"uninger}, \citenamefont {Nunner}, \citenamefont {Knetter},
  \citenamefont {Schmidt}, \citenamefont {Uhrig}, \citenamefont {Kopp},
  \citenamefont {Freimuth}, \citenamefont {Ammerahl}, \citenamefont
  {B\"uchner},\ and\ \citenamefont {Revcolevschi}}]{Windt2001}%
  \BibitemOpen
  \bibfield  {author} {\bibinfo {author} {\bibfnamefont {M.}~\bibnamefont
  {Windt}}, \bibinfo {author} {\bibfnamefont {M.}~\bibnamefont {Gr\"uninger}},
  \bibinfo {author} {\bibfnamefont {T.}~\bibnamefont {Nunner}}, \bibinfo
  {author} {\bibfnamefont {C.}~\bibnamefont {Knetter}}, \bibinfo {author}
  {\bibfnamefont {K.~P.}\ \bibnamefont {Schmidt}}, \bibinfo {author}
  {\bibfnamefont {G.~S.}\ \bibnamefont {Uhrig}}, \bibinfo {author}
  {\bibfnamefont {T.}~\bibnamefont {Kopp}}, \bibinfo {author} {\bibfnamefont
  {A.}~\bibnamefont {Freimuth}}, \bibinfo {author} {\bibfnamefont
  {U.}~\bibnamefont {Ammerahl}}, \bibinfo {author} {\bibfnamefont
  {B.}~\bibnamefont {B\"uchner}}, \ and\ \bibinfo {author} {\bibfnamefont
  {A.}~\bibnamefont {Revcolevschi}},\ }\href@noop {} {\bibfield  {journal}
  {\bibinfo  {journal} {Phys. Rev. Lett.}\ }\textbf {\bibinfo {volume} {87}},\
  \bibinfo {pages} {127002} (\bibinfo {year} {2001})}\BibitemShut {NoStop}%
\bibitem [{\citenamefont {Martin}\ and\ \citenamefont
  {Illas}(1997)}]{Martin1997}%
  \BibitemOpen
  \bibfield  {author} {\bibinfo {author} {\bibfnamefont {R.~L.}\ \bibnamefont
  {Martin}}\ and\ \bibinfo {author} {\bibfnamefont {F.}~\bibnamefont {Illas}},\
  }\href@noop {} {\bibfield  {journal} {\bibinfo  {journal} {Phys. Rev. Lett.}\
  }\textbf {\bibinfo {volume} {79}},\ \bibinfo {pages} {1539--1542} (\bibinfo
  {year} {1997})}\BibitemShut {NoStop}%
\bibitem [{\citenamefont {Sa\'ul}\ and\ \citenamefont
  {Radtke}(2014)}]{Saul2014}%
  \BibitemOpen
  \bibfield  {author} {\bibinfo {author} {\bibfnamefont {A.}~\bibnamefont
  {Sa\'ul}}\ and\ \bibinfo {author} {\bibfnamefont {G.}~\bibnamefont
  {Radtke}},\ }\href@noop {} {\bibfield  {journal} {\bibinfo  {journal} {Phys.
  Rev. B}\ }\textbf {\bibinfo {volume} {89}},\ \bibinfo {pages} {104414}
  (\bibinfo {year} {2014})}\BibitemShut {NoStop}%
\bibitem [{\citenamefont {Jeschke}\ \emph {et~al.}(2011)\citenamefont
  {Jeschke}, \citenamefont {Opahle}, \citenamefont {Kandpal}, \citenamefont
  {Valent\'{\i}}, \citenamefont {Das}, \citenamefont {Saha-Dasgupta},
  \citenamefont {Janson}, \citenamefont {Rosner}, \citenamefont {Br\"uhl},
  \citenamefont {Wolf}, \citenamefont {Lang}, \citenamefont {Richter},
  \citenamefont {Hu}, \citenamefont {Wang}, \citenamefont {Peters},
  \citenamefont {Pruschke},\ and\ \citenamefont {Honecker}}]{Jeschke2011}%
  \BibitemOpen
  \bibfield  {author} {\bibinfo {author} {\bibfnamefont {H.}~\bibnamefont
  {Jeschke}}, \bibinfo {author} {\bibfnamefont {I.}~\bibnamefont {Opahle}},
  \bibinfo {author} {\bibfnamefont {H.}~\bibnamefont {Kandpal}}, \bibinfo
  {author} {\bibfnamefont {R.}~\bibnamefont {Valent\'{\i}}}, \bibinfo {author}
  {\bibfnamefont {H.}~\bibnamefont {Das}}, \bibinfo {author} {\bibfnamefont
  {T.}~\bibnamefont {Saha-Dasgupta}}, \bibinfo {author} {\bibfnamefont
  {O.}~\bibnamefont {Janson}}, \bibinfo {author} {\bibfnamefont
  {H.}~\bibnamefont {Rosner}}, \bibinfo {author} {\bibfnamefont
  {A.}~\bibnamefont {Br\"uhl}}, \bibinfo {author} {\bibfnamefont
  {B.}~\bibnamefont {Wolf}}, \bibinfo {author} {\bibfnamefont {M.}~\bibnamefont
  {Lang}}, \bibinfo {author} {\bibfnamefont {J.}~\bibnamefont {Richter}},
  \bibinfo {author} {\bibfnamefont {S.}~\bibnamefont {Hu}}, \bibinfo {author}
  {\bibfnamefont {X.}~\bibnamefont {Wang}}, \bibinfo {author} {\bibfnamefont
  {R.}~\bibnamefont {Peters}}, \bibinfo {author} {\bibfnamefont
  {T.}~\bibnamefont {Pruschke}}, \ and\ \bibinfo {author} {\bibfnamefont
  {A.}~\bibnamefont {Honecker}},\ }\href@noop {} {\bibfield  {journal}
  {\bibinfo  {journal} {Phys. Rev. Lett.}\ }\textbf {\bibinfo {volume} {106}},\
  \bibinfo {pages} {217201} (\bibinfo {year} {2011})}\BibitemShut {NoStop}%
\bibitem [{\citenamefont {Zhitomirsky}(2006)}]{Zhitomirsky2006}%
  \BibitemOpen
  \bibfield  {author} {\bibinfo {author} {\bibfnamefont {M.~E.}\ \bibnamefont
  {Zhitomirsky}},\ }\href@noop {} {\bibfield  {journal} {\bibinfo  {journal}
  {Phys. Rev. B}\ }\textbf {\bibinfo {volume} {73}},\ \bibinfo {pages} {100404}
  (\bibinfo {year} {2006})}\BibitemShut {NoStop}%
\bibitem [{\citenamefont {Zhitomirsky}\ and\ \citenamefont
  {Chernyshev}(2013)}]{Zhitomirsky2013}%
  \BibitemOpen
  \bibfield  {author} {\bibinfo {author} {\bibfnamefont {M.~E.}\ \bibnamefont
  {Zhitomirsky}}\ and\ \bibinfo {author} {\bibfnamefont {A.~L.}\ \bibnamefont
  {Chernyshev}},\ }\href@noop {} {\bibfield  {journal} {\bibinfo  {journal}
  {Rev. Mod. Phys.}\ }\textbf {\bibinfo {volume} {85}},\ \bibinfo {pages}
  {219--242} (\bibinfo {year} {2013})}\BibitemShut {NoStop}%
\bibitem [{\citenamefont {Lorenzana}\ and\ \citenamefont
  {Sawatzky}(1995{\natexlab{a}})}]{Lorenzana1995a}%
  \BibitemOpen
  \bibfield  {author} {\bibinfo {author} {\bibfnamefont {J.}~\bibnamefont
  {Lorenzana}}\ and\ \bibinfo {author} {\bibfnamefont {G.~A.}\ \bibnamefont
  {Sawatzky}},\ }\href@noop {} {\bibfield  {journal} {\bibinfo  {journal}
  {Phys. Rev. Lett.}\ }\textbf {\bibinfo {volume} {74}},\ \bibinfo {pages}
  {1867--1870} (\bibinfo {year} {1995}{\natexlab{a}})}\BibitemShut {NoStop}%
\bibitem [{\citenamefont {Lorenzana}\ and\ \citenamefont
  {Sawatzky}(1995{\natexlab{b}})}]{Lorenzana1995b}%
  \BibitemOpen
  \bibfield  {author} {\bibinfo {author} {\bibfnamefont {J.}~\bibnamefont
  {Lorenzana}}\ and\ \bibinfo {author} {\bibfnamefont {G.~A.}\ \bibnamefont
  {Sawatzky}},\ }\href@noop {} {\bibfield  {journal} {\bibinfo  {journal}
  {Phys. Rev. B}\ }\textbf {\bibinfo {volume} {52}},\ \bibinfo {pages}
  {9576--9589} (\bibinfo {year} {1995}{\natexlab{b}})}\BibitemShut {NoStop}%
\bibitem [{\citenamefont {Lorenzana}\ and\ \citenamefont
  {Eder}(1997)}]{Lorenzana1997}%
  \BibitemOpen
  \bibfield  {author} {\bibinfo {author} {\bibfnamefont {J.}~\bibnamefont
  {Lorenzana}}\ and\ \bibinfo {author} {\bibfnamefont {R.}~\bibnamefont
  {Eder}},\ }\href@noop {} {\bibfield  {journal} {\bibinfo  {journal} {Phys.
  Rev. B}\ }\textbf {\bibinfo {volume} {55}},\ \bibinfo {pages} {R3358--R3361}
  (\bibinfo {year} {1997})}\BibitemShut {NoStop}%
\bibitem [{\citenamefont {Suzuura}\ \emph {et~al.}(1996)\citenamefont
  {Suzuura}, \citenamefont {Yasuhara}, \citenamefont {Furusaki}, \citenamefont
  {Nagaosa},\ and\ \citenamefont {Tokura}}]{Suzuura1996}%
  \BibitemOpen
  \bibfield  {author} {\bibinfo {author} {\bibfnamefont {H.}~\bibnamefont
  {Suzuura}}, \bibinfo {author} {\bibfnamefont {H.}~\bibnamefont {Yasuhara}},
  \bibinfo {author} {\bibfnamefont {A.}~\bibnamefont {Furusaki}}, \bibinfo
  {author} {\bibfnamefont {N.}~\bibnamefont {Nagaosa}}, \ and\ \bibinfo
  {author} {\bibfnamefont {Y.}~\bibnamefont {Tokura}},\ }\href@noop {}
  {\bibfield  {journal} {\bibinfo  {journal} {Phys. Rev. Lett.}\ }\textbf
  {\bibinfo {volume} {76}},\ \bibinfo {pages} {2579--2582} (\bibinfo {year}
  {1996})}\BibitemShut {NoStop}%
\bibitem [{\citenamefont {{R\~o\~om}}\ \emph
  {et~al.}(2004{\natexlab{b}})\citenamefont {{R\~o\~om}}, \citenamefont
  {H\"uvonen}, \citenamefont {Nagel}, \citenamefont {Wang},\ and\ \citenamefont
  {Kremer}}]{Room2004a}%
  \BibitemOpen
  \bibfield  {author} {\bibinfo {author} {\bibfnamefont {T.}~\bibnamefont
  {{R\~o\~om}}}, \bibinfo {author} {\bibfnamefont {D.}~\bibnamefont
  {H\"uvonen}}, \bibinfo {author} {\bibfnamefont {U.}~\bibnamefont {Nagel}},
  \bibinfo {author} {\bibfnamefont {Y.-J.}\ \bibnamefont {Wang}}, \ and\
  \bibinfo {author} {\bibfnamefont {R.~K.}\ \bibnamefont {Kremer}},\
  }\href@noop {} {\bibfield  {journal} {\bibinfo  {journal} {Phys. Rev. B}\
  }\textbf {\bibinfo {volume} {69}},\ \bibinfo {pages} {144410} (\bibinfo
  {year} {2004}{\natexlab{b}})}\BibitemShut {NoStop}%
\bibitem [{\citenamefont {C\'epas}\ and\ \citenamefont
  {Ziman}(2004)}]{Cepas2004}%
  \BibitemOpen
  \bibfield  {author} {\bibinfo {author} {\bibfnamefont {O.}~\bibnamefont
  {C\'epas}}\ and\ \bibinfo {author} {\bibfnamefont {T.}~\bibnamefont
  {Ziman}},\ }\href@noop {} {\bibfield  {journal} {\bibinfo  {journal} {Phys.
  Rev. B}\ }\textbf {\bibinfo {volume} {70}},\ \bibinfo {pages} {024404}
  (\bibinfo {year} {2004})}\BibitemShut {NoStop}%
\bibitem [{\citenamefont {Baym}(1969)}]{Baym}%
  \BibitemOpen
  \bibfield  {author} {\bibinfo {author} {\bibfnamefont {G.}~\bibnamefont
  {Baym}},\ }\href@noop {} {\emph {\bibinfo {title} {Lectures on Quantum
  Mechanics}}}\ (\bibinfo  {publisher} {Westview Press},\ \bibinfo {year}
  {1969})\BibitemShut {NoStop}%
\bibitem [{\citenamefont {Kogut}(1979)}]{Kogut79}%
  \BibitemOpen
  \bibfield  {author} {\bibinfo {author} {\bibfnamefont {J.~B.}\ \bibnamefont
  {Kogut}},\ }\href@noop {} {\bibfield  {journal} {\bibinfo  {journal} {Rev.
  Mod. Phys.}\ }\textbf {\bibinfo {volume} {51}},\ \bibinfo {pages} {659--713}
  (\bibinfo {year} {1979})}\BibitemShut {NoStop}%
\bibitem [{\citenamefont {Cabra}\ \emph {et~al.}(1998)\citenamefont {Cabra},
  \citenamefont {Honecker},\ and\ \citenamefont {Pujol}}]{CHP98}%
  \BibitemOpen
  \bibfield  {author} {\bibinfo {author} {\bibfnamefont {D.~C.}\ \bibnamefont
  {Cabra}}, \bibinfo {author} {\bibfnamefont {A.}~\bibnamefont {Honecker}}, \
  and\ \bibinfo {author} {\bibfnamefont {P.}~\bibnamefont {Pujol}},\
  }\href@noop {} {\bibfield  {journal} {\bibinfo  {journal} {Phys. Rev. B}\
  }\textbf {\bibinfo {volume} {58}},\ \bibinfo {pages} {6241--6257} (\bibinfo
  {year} {1998})}\BibitemShut {NoStop}%
\bibitem [{\citenamefont {Honecker}(1999)}]{perturb99}%
  \BibitemOpen
  \bibfield  {author} {\bibinfo {author} {\bibfnamefont {A.}~\bibnamefont
  {Honecker}},\ }\href@noop {} {\bibfield  {journal} {\bibinfo  {journal}
  {Phys. Rev. B}\ }\textbf {\bibinfo {volume} {59}},\ \bibinfo {pages}
  {6790--6794} (\bibinfo {year} {1999})}\BibitemShut {NoStop}%
\bibitem [{Sup()}]{SuppMat}%
  \BibitemOpen
  \href@noop {} {}\bibinfo {note} {See ancillary files for the expression of
  the effective Hamiltonian up to the fifth order and the ground state energy
  up to the seventh order in the form of a Mathematica notebook, PDF, or plain
  text.}\BibitemShut {Stop}%
\bibitem [{\citenamefont {Reigrotzki}\ \emph {et~al.}(1994)\citenamefont
  {Reigrotzki}, \citenamefont {Tsunetsugu},\ and\ \citenamefont
  {Rice}}]{Reigrotzki1994}%
  \BibitemOpen
  \bibfield  {author} {\bibinfo {author} {\bibfnamefont {M.}~\bibnamefont
  {Reigrotzki}}, \bibinfo {author} {\bibfnamefont {H.}~\bibnamefont
  {Tsunetsugu}}, \ and\ \bibinfo {author} {\bibfnamefont {T.~M.}\ \bibnamefont
  {Rice}},\ }\href@noop {} {\bibfield  {journal} {\bibinfo  {journal} {J.
  Phys.: Condens. Matter}\ }\textbf {\bibinfo {volume} {6}},\ \bibinfo {pages}
  {9235--9245} (\bibinfo {year} {1994})}\BibitemShut {NoStop}%
\bibitem [{\citenamefont {Knetter}\ and\ \citenamefont
  {Uhrig}(2000)}]{Knetter2000}%
  \BibitemOpen
  \bibfield  {author} {\bibinfo {author} {\bibfnamefont {C.}~\bibnamefont
  {Knetter}}\ and\ \bibinfo {author} {\bibfnamefont {G.~S.}\ \bibnamefont
  {Uhrig}},\ }\href@noop {} {\bibfield  {journal} {\bibinfo  {journal} {Eur.
  Phys. J. B}\ }\textbf {\bibinfo {volume} {13}},\ \bibinfo {pages} {209--225}
  (\bibinfo {year} {2000})}\BibitemShut {NoStop}%
\bibitem [{\citenamefont {Barnes}\ \emph {et~al.}(1999)\citenamefont {Barnes},
  \citenamefont {Riera},\ and\ \citenamefont {Tennant}}]{Barnes1999}%
  \BibitemOpen
  \bibfield  {author} {\bibinfo {author} {\bibfnamefont {T.}~\bibnamefont
  {Barnes}}, \bibinfo {author} {\bibfnamefont {J.}~\bibnamefont {Riera}}, \
  and\ \bibinfo {author} {\bibfnamefont {D.~A.}\ \bibnamefont {Tennant}},\
  }\href@noop {} {\bibfield  {journal} {\bibinfo  {journal} {Phys. Rev. B}\
  }\textbf {\bibinfo {volume} {59}},\ \bibinfo {pages} {11384--11397} (\bibinfo
  {year} {1999})}\BibitemShut {NoStop}%
\end{thebibliography}%

\end{document}